\documentclass[twocolumn,floatfix]{aastex701}
\usepackage{hyperref}

\usepackage{amsmath}
\let\cleardoublepage=\clearpage

\begin{document}

\title{Radio and X-ray Observations of the Transitional Supernova 2019yvr:  Insights into the Progenitor Mass-Loss History}
\shorttitle{Radio and X-ray Insights into SN 2019yvr}
\shortauthors{Baer-way et al.}

\newcommand{\UCB}{\affiliation{Department of Astronomy, University of California, Berkeley, CA 94720-3411, USA}}
\author[0009-0004-7268-7283]{Raphael Baer-way}
\affil{Department of Astronomy, University of Virginia, 530 McCormick Rd, Charlottesville, VA 22904, USA}
\affiliation{National Radio Astronomy Observatory,
520 Edgemont Rd, Charlottesville VA 22903, USA}
\email{placeholder@gmail.com}
\author[0000-0002-0844-6563]{Poonam Chandra}
\affiliation{National Radio Astronomy Observatory,
520 Edgemont Rd, Charlottesville VA 22903, USA}
\email{placeholder@gmail.com}
\author[0000-0001-7132-0333]{Maryam Modjaz}
\affil{Department of Astronomy, University of Virginia, 530 McCormick Rd, Charlottesville, VA 22904, USA}
\email{placeholder@email.com}
\author[0000-0002-8070-5400]{Nayana A. J.}
\affiliation{Berkeley Center for Multi-messenger Research on Astrophysical Transients and Outreach (Multi-RAPTOR), University of California, Berkeley, CA 94720-3411, USA}
\email{placeholder@gmail.com}
\author[0000-0003-2611-7269]{Keiichi Maeda}
\affiliation{Department of Astronomy, Kyoto University, Kitashirakawa-Oiwake-cho, Sakyo-ku, Kyoto, 606-8502. Japan}
\email{keiichi.maeda@kusastro.kyoto-u.ac.jp}  
\author[0000-0002-4449-9152]{Katie Auchettl}
\affiliation{Department of Astronomy and Astrophysics, University of California, Santa Cruz, CA 95064, USA}
\affiliation{OzGrav, School of Physics, The University of Melbourne, Parkville, VIC, Australia}
\email{placeholder@gmail.com}
\author[0000-0001-7081-0082]{Maria R. Drout}
\affiliation{David A. Dunlap Department of Astronomy \& Astrophysics, University of Toronto, 50 St George St, Toronto, ON M5S 3H4, Canada}
\email{placeholder@gmail.com}
\author[0000-0002-5740-7747]{Charles~D.~Kilpatrick}
\affiliation{Center for Interdisciplinary Exploration and Research in Astrophysics (CIERA), Northwestern University, Evanston, IL 60208, USA}
\email{ckilpatrick@northwestern.edu}
\email{placeholder@gmail.com}
\author[0000-0003-2404-0018]{Alak K. Ray}
\affiliation{Homi Bhabha Centre for Science Education, TIFR, Mumbai 400088, India}
\email{placeholder@gmail.com}

\author[0000-0003-4501-8100]{Stuart D. Ryder}
\affiliation{School of Mathematical and Physical Sciences, Macquarie University, Sydney, NSW 2109, Australia}
\affiliation{Astrophysics and Space Technologies Research Centre, Macquarie University, Sydney, NSW 2109, Australia}
\email{placeholder@gmail.com}



\begin{abstract}
The final life stages of the massive star progenitors of stripped-envelope supernovae (SESNe) are still an open question, especially when it comes to the timing and magnitude of the progenitor stripping. Observing SESNe across the electromagnetic spectrum allows for the most direct constraints on mass loss in the final stages of progenitor evolution. In this work, we present radio (GMRT+VLA) and X-ray (Swift+Chandra) observations of SN 2019yvr obtained from 18-1786 days post-explosion. SN 2019yvr was a type Ib supernova (SN Ib, with strong helium but little or no hydrogen features) that transitioned into a type IIn supernova (SN IIn, with shock-driven hydrogen features from interaction with circumstellar material (CSM)) at $\sim$ 100 days post-explosion. The radio evolution is best-fit by a synchrotron self-absorbed model with a $\rho \propto r^{-1.54 \pm 0.31}$ CSM density profile, suggesting a decreasing mass-loss rate from the progenitor in the years leading up to the explosion. The radio-derived shock speed is high, more than 30,000 km/s at early times, suggesting a compact progenitor star. The combined radio and X-ray data probe CSM that extends from  $1.7 \times 10^{16}$ cm up to $\sim$ 20$\times10^{16}$ cm and was created by mass loss from $\sim (1-6) \times10^{-5} \rm{M_{\odot} yr^{-1}} $ (assuming a CSM speed of 100 km/s). The combined dataset rules out any \emph{dramatic} jump in CSM density (which was seen in the optical analog SN 2014C) associated with the emergence of optical hydrogen emission lines. We place SN 2019yvr in context with similar transitional SNe and discuss implications for the progenitor.

\end{abstract}


\section{Introduction} 
\label{sec:intro}
The explosions of massive stars in core-collapse supernovae (CCSNe) are some of the most energetic events in our universe.
Certain massive stars explode as ``stripped" SNe, where there is very little or no hydrogen or helium in the optical spectra of the exploding star \citep{Uomoto_Kirshner,Harkness_wheeler1990,Clocchiatti,Filippenko_1997,Modjaz_2019}. These stripped-envelope supernovae (SESNe) make up roughly $30\%$ of all observed CCSNe by volumetric rate \citep{Shivvers_2017,Perley_2020}, with these objects being further classified into Type IIb (some hydrogen at early times), Type Ib (no strong hydrogen), and Type Ic (no strong hydrogen+helium), although details on the true presence of hydrogen or helium at certain phases are unclear (see e.g. \cite{Williamson_2021,Yesmin_2025,Lu_2026}). Nevertheless, it is clear that SNe Ib must lose most if not all of their hydrogen envelope before the explosion \citep{Dessart_2020,Dong_2024}. 

While it was initially thought that the envelope was stripped by strong winds in Wolf-Rayet (WR) stars \citep{Woosley_1993,Heger_2003,Langer_2012,Groh_2013,Aguilera_2023}, the aforementioned large fraction of SESNe found in volumetric surveys made it clear that not all of these explosions can come from such massive WR stars $> 30 \rm M_{\odot}$ \citep{Smith_2011}. Recent work has suggested that perhaps all SESNe can be explained by binary stripping of lower-mass stars ($< 20 M_{\odot}$) \citep{Phillip_92,Nomoto_1995,Phillip_2004,Dessart_2020,Woosley_2021,Ercolino_2025}, where the companion strips the star at some point in the years preceding the explosion, and the star explodes thereafter. However, others have argued that at least some SESNe must still be the explosions of Wolf-Rayet stars \citep{Gal-Yam_2024}. 

The plausibility of binary stripping is confirmed by the fact that mass transfer can occur before or during the helium star phase \citep{Tsai_2026}. In any case, questions still persist around when exactly the mass transfer or stripping occurs and when the explosion happens afterwards (see e.g. \citet{Fang_2019}).  A diversity in mass-loss timescales would not be surprising for binary progenitors, as binary interaction should create a large diversity in circumstellar environments depending on the type of mass transfer and initial binary separation \citep{Scherbak_2025,Dessart_2020,Brethauer_2022,Mandal_2026}.

In certain supernovae (SNe), it is clear that at least some of the stripped material (no matter the mechanism) stays relatively close to the SN (meaning it was likely only lost in the final decades pre-explosion) \citep{Chevalier_2010,Brethauer_2022}. At optical wavelengths, there have been certain cases where an object that was initially a SESN unambiguously encountered hydrogen-rich circumstellar material (CSM) formed through extreme mass-loss,  such as SN~2014C \citep{Danny_2014C}. SN 2014C thus evolved into a SN IIn, which experiences the highest rates of hydrogen-rich mass-loss of all SNe \citep{Weiler_1990,Margutti_2017,Smith_2016,Chandra_2020,Hillenkamp_2026}. Multiwavelength follow-up of these transition objects and SNe IIn often reveals highly complex CSM with multi-stage mass-loss histories \citep{Brethauer_2022,Thomas_2022,Sfaradi_2024}. The object we analyze in this work, SN 2019yvr experienced a similar metamorphosis at optical wavelengths: it transitioned from a SN Ib to a SN IIn at $\sim$ 100 days post-explosion \citep{Kilpatrick_2021,Sun_2021,Ferrari_2024}. We use five years of Giant Metrewave Radio Telescope (GMRT), Karl Jansky Very Large Array (VLA)  radio data and \textit{Chandra} and \textit{Swift} X-ray data to explore SN~2019yvr's CSM environment.

In cases where a SESN is surrounded by dense CSM, radio and X-ray emission is produced by the forward and reverse shocks generated by ejecta-CSM interaction. The forward shock produces radio synchrotron and thermal or nonthermal X-ray emission \citep{Chevalier_17}, while the reverse shock is expected to produce X-ray emission (usually thermal in SESNe \cite{Dwarkadas_2025}). In SESNe, the CSM is significantly less dense than in the case of SNe IIn, and thus the X-ray emission is often much less luminous as the thermal component is weaker \citep{Dwarkadas_2012}. 

Mass-loss rates measured from typical SESNe range from $10^{-3}-10^{-6} \rm M_{\odot}\rm yr^{-1}$ \citep{MAeda_2021,Bietenholz_2021,Sfaradi_2025}, with the speed of the unshocked material typically assumed at $\sim$ 1000 km/s for compact type Ib/Ic progenitors based on the escape speed/wind velocity measured for Wolf-Rayet stars in our own galaxy \citep{2007_AIP,Grafener_2017,Chandra_2020}. The speed of the unshocked CSM can only be measured through high-resolution optical spectra, and has only been measured directly in certain rare SESNe such as SNe Ibn (where speeds are measured from 1000-2000 km/s) and the previously discussed SN Ib 2014C with a CSM speed around 100 km/s \citep{Frannson_96,Danny_2014C,Dong_2024,RBW_2025}. The inferred mass-loss rates are consistent with strong winds from Wolf-Rayet stars \citep{Woosley_1993}. However, these rates are indeed also consistent with expected mass transfer rates from binary stripping \citep{Phillip_92,Nomoto_1995,Dessart_2020,Ercolino_2025}. 

The CSM interaction itself can also reveal details on the progenitor regardless of mass-loss rate by giving clues as to the extended or compact nature of the star that exploded through the shock speed measured at radio wavelengths \citep{Chevalier_2010,Ouchi_2017}. A further diagnostic for the progenitor is understanding the duration of the mass loss. Following up SESNe out to late times (as the ejecta continue to catch up to the slower-moving CSM) allows for deep constraints on mass loss centuries before the explosion, which can make it clear whether the star experienced wind-like constant mass loss or sudden upticks more characteristic of complex mass-loss mechanisms like binary inspiral, pulsations or  gravity waves \citep{Quataert_2012,Ouchi_2017,Laplace_2025,Goldberg_2025,Sengupta_2026}.

In this work, we seek to characterize the CSM around the SN Ib 2019yvr. SN~2019yvr is a SN that has been observed extensively at optical wavelengths \citep{Kilpatrick_2021,Sun_2021,Ferrari_2024} due to the changes in its optical spectra from hydrogen-poor or helium-rich (type Ib) to hydrogen-rich with interaction-driven emission lines (type IIn). Additionally, the SN had pre-explosion HST imaging that revealed a progenitor candidate \citep{Kilpatrick_2021,Sun_2021}. SED modeling of the progenitor candidate suggested an extended supergiant (radius of $320_{-50}^{+30} \rm R_{\odot}$) for the progenitor \citep{Kilpatrick_2021}, but later work suggested that the source at the progenitor location is an amalgamation of two stars, with the supergiant being the binary companion and the progenitor itself being compact \citep{Sun_2021} with a radius $\sim$ 20 $R_{\rm \odot}$ (with the compactness defined relative to a supergiant progenitor). 

The spectra developed clear hydrogen features between 80--120 days post-explosion, leading to the implication that the SN ejecta had begun interacting with dense CSM \citep{Ferrari_2024}. We seek to draw conclusions based on an extensive radio and X-ray dataset and examine the results mentioned above. The SN was discovered in NGC 4666 on December 27, 2019, with the explosion date constrained to December 22, 2019 based on the last non-detection \citep{Ferrari_2024}. The distance to the SN is constrained to 14.7 $\pm$ 1.5 Mpc from a previous SN discovered in this galaxy  \citep{Shappee_2016}. We use these values for the date of explosion and distance to the SN throughout this work. In Section \ref{sec:Data}, we lay out the data we obtained and the data reductions. In Section \ref{sec:Analysis}, we elaborate on the modeling we performed of the data. In Section \ref{sec:Disc} and \ref{sec:Conc}, we discuss the implications of our results and conclude.
\section{Data}\label{sec:Data}

We obtained 5 years of radio data on SN 2019yvr with the GMRT and also used data obtained from 2020-2024 with the VLA. We also report on the \textit{Chandra} X-ray detection of the SN as well as report the data obtained with the X-ray telescope onboard \textit{Swift} (\textit{Swift-XRT}).
\begin{figure*}[htb!]
\gridline{
  \parbox{0.35\textwidth}{\centering Swift-XRT 0.2-10 keV image 03/2020\\ \includegraphics[width=0.35\textwidth]{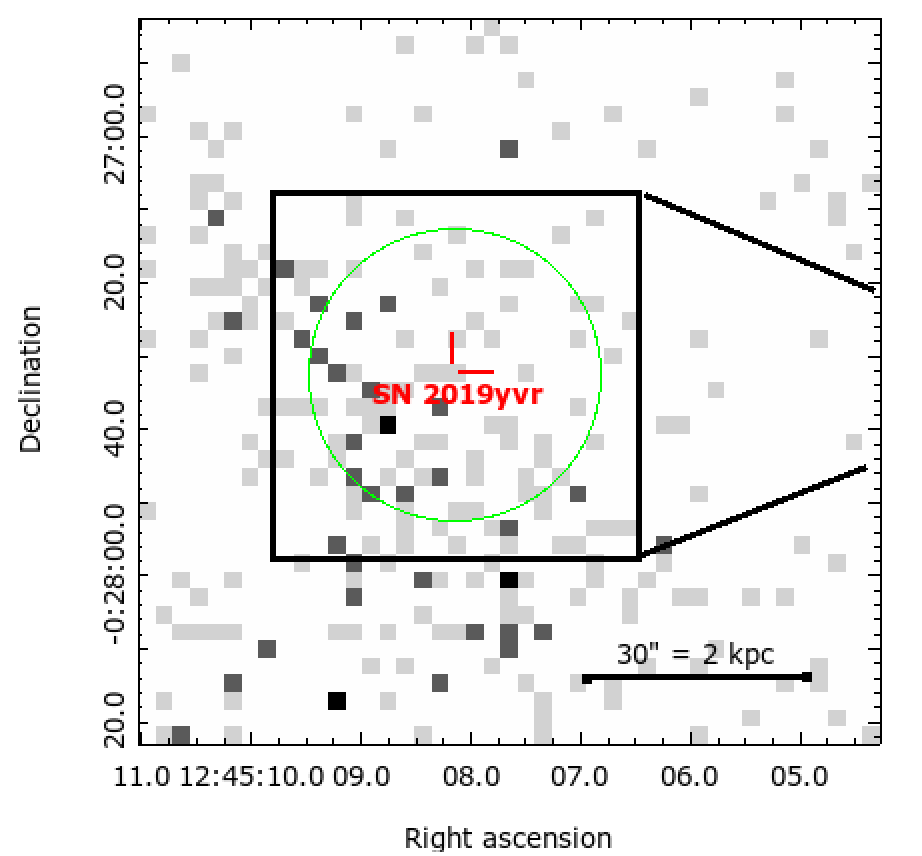}}\hspace{-2.0cm}
  \parbox{0.36\textwidth}{\centering \textit{Chandra} 0.3-10 keV image 01/2020\\ \includegraphics[width=0.36\textwidth]{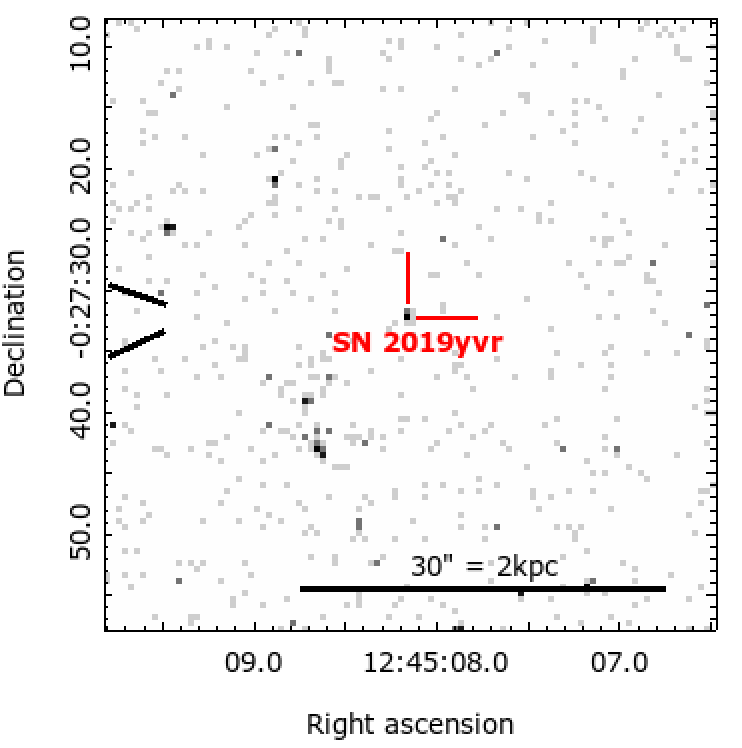}}
}
\vspace{-0.5cm}
\gridline{
  \parbox{0.36\textwidth}{\centering GMRT 1.26 GHz image 08/2022\\ \includegraphics[width=0.36\textwidth]{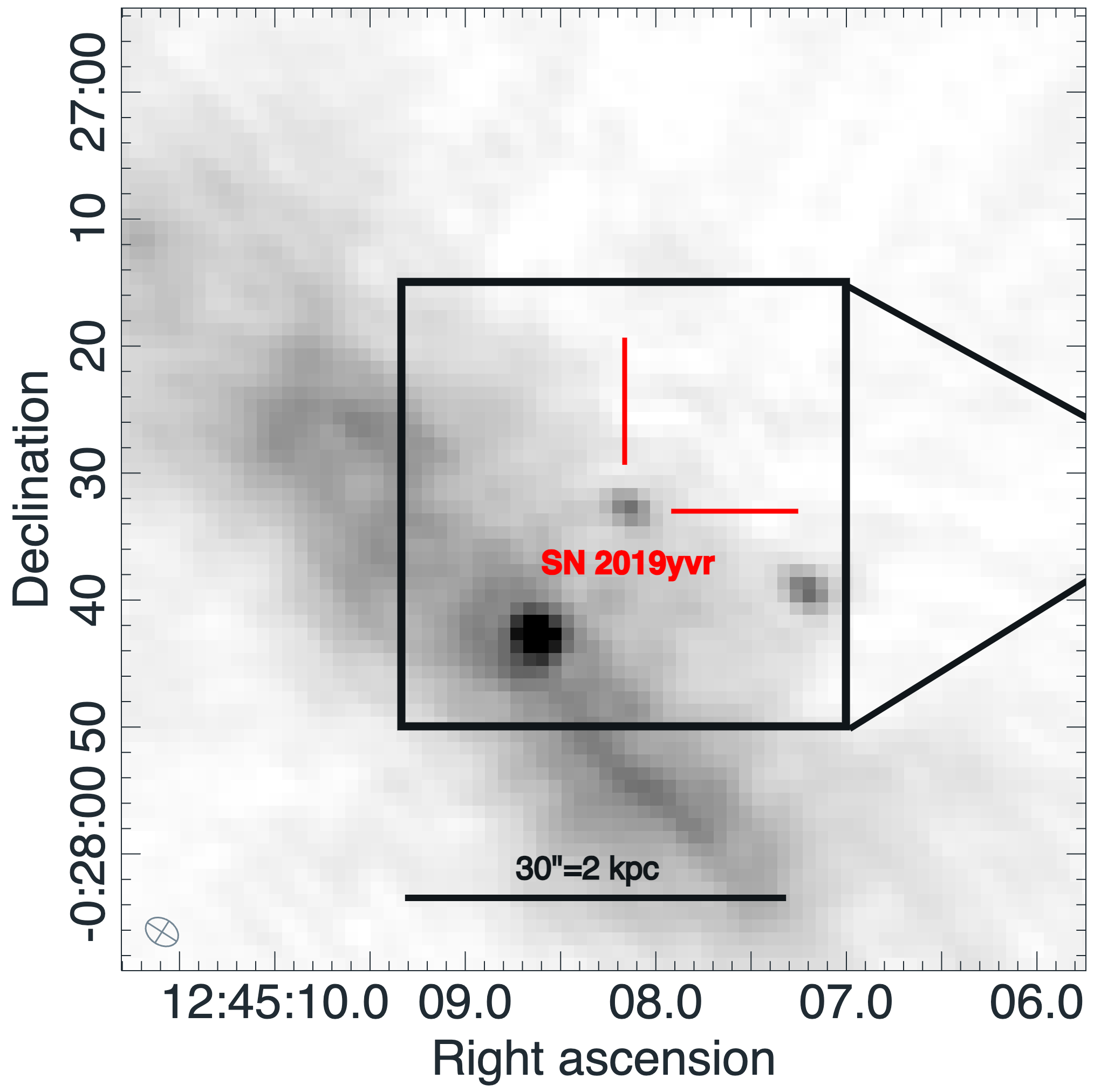}}\hspace{-2.0cm}
  \parbox{0.35\textwidth}{\centering VLA 4.93 GHz image 11/2024\\ \includegraphics[width=0.35\textwidth]{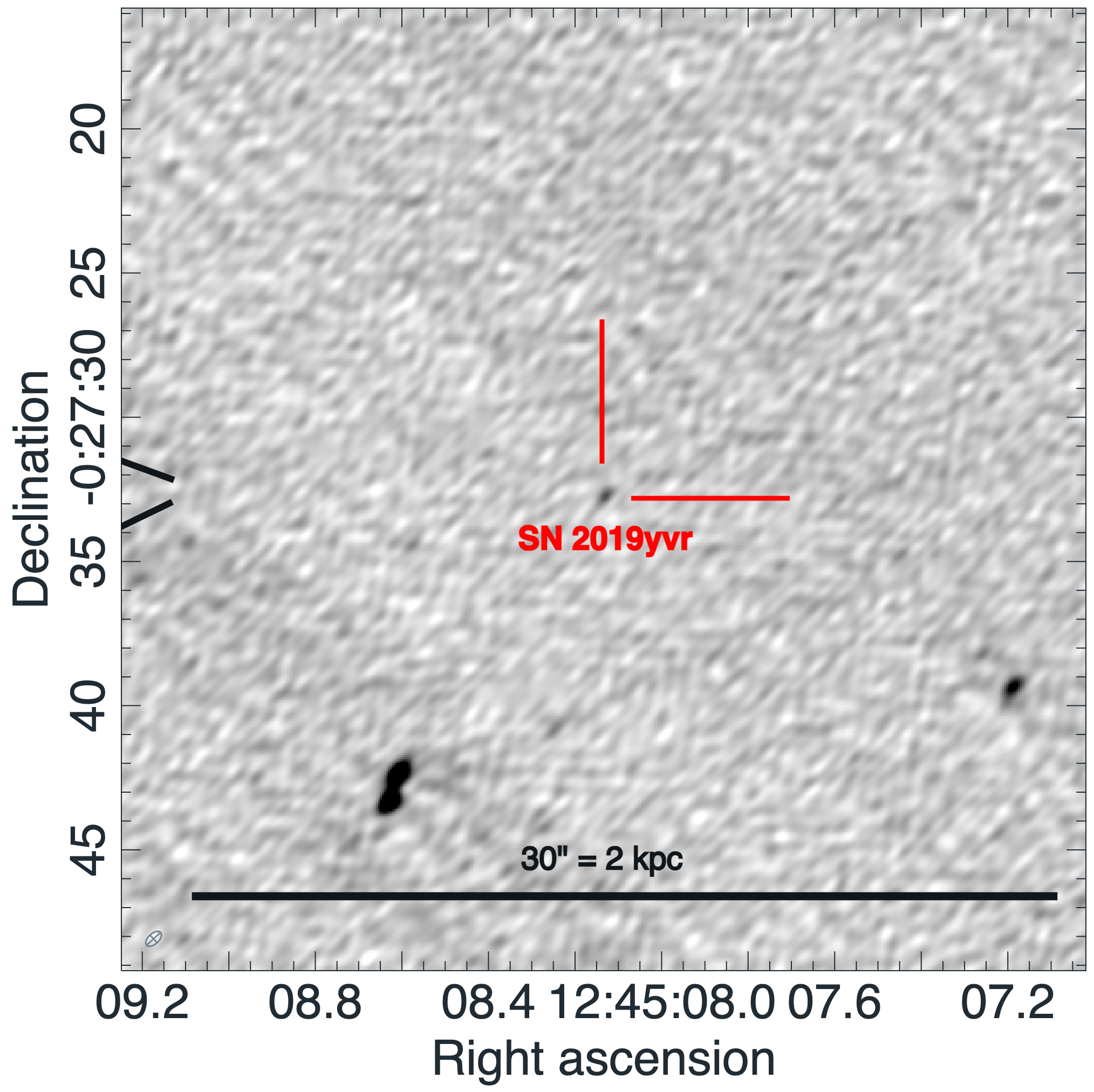}}
}

\caption{Radio and X-ray images of SN 2019yvr at various points in its evolution. The GMRT image has $\sim$ 2$''$ resolution, while the VLA image has $\sim$ 0.4$''$ resolution. We show the 20$''$ region around the SN in the \textit{Swift} image that was used to extract \textit{Swift} pre- and post-explosion X-ray spectra. We also show the boxed region within the larger FOV images (\textit{Swift}, GMRT) that the higher-resolution image encompasses (\textit{Chandra}, VLA). The nucleus of NGC 4666 is visible in the GMRT and VLA images, 15 arcsec to the SE of SN 2019yvr.}
\label{fig:Overall_images}
\end{figure*}
\subsection{Radio GMRT and VLA Observations}
We took 11 epochs of GMRT data of SN 2019yvr largely as part of the series of GMRT proposals under the GLIMPSE program (GMRT Low-frequency Investigation of Massive Progenitors of SNe Explosions -- PI P.Chandra). All GMRT epochs consisted of at least one band 5 observation (1000--1450\,MHz), and almost all epochs also had band 4 observations (550--900\,MHz). Four of the later epochs also had band 3 data (250--500\,MHz). GMRT band 5 data were taken with 400 MHz bandwidth centered at 1.25 GHz, with 2048 channels. GMRT band 4 data were taken with 200 or 400 MHz bandwidth centered at 0.65 GHz or 0.75 GHz, with 2048 channels. GMRT band 3 data were taken with 200 MHz bandwidth centered at 0.4 GHz, with 2048 channels.  

All GMRT data were reduced using the CAPTURE pipeline \citep{GMRT_pipeline}, with multiple rounds of phase-only self-calibration performed within the pipeline. The host galaxy is at $\sim 15$ arcsec away, and is also radio bright (at $\sim$ 10 mJy at 0.75 GHz), but the high resolution of the GMRT at band 5  ($\sim$ 1 arcsecond) allowed for relatively straightforward fitting using \textit{imfit} to measure the flux density of the SN. At later epochs, the SN was not detected in band 5. We report 3$\sigma$ upper limits based on the image RMS in a region centered around the SN that is multiple times the beam size. 

At bands 3 and 4, however, the data reduction proved more difficult due to contamination from the host. At band 4, we are able to extract the SN flux at certain epochs following careful flagging after self-calibration. However, the host galaxy and the changing beam orientation as the object is observed at different times made certain data unusable due to contamination. We only report upper limits in band 3 due to flux from the host galaxy preventing the recovery of the uncontaminated SN flux.

There were 5 epochs of VLA observations of 2019yvr from L--K bands (1--26.5\,GHz) from proposals 20A-571 (PI K.Auchettl) and 22A-204 (PI M.Drout) along with three serendipitous  S and C band detections in radio imaging of NGC 4666 from proposals 20A-018 (PI Y.Stein) and 24B-453 (PI M.Gorski). The full dataset with details is reported in Table \ref{tab:radio_obs}.  Reductions for almost all VLA data were straightforward, as the majority of the data were taken in high-resolution VLA configurations (configuration A or B). The phase calibrator used was J1246-0730, and 3C286 was used as the bandpass and flux calibrator. We used the VLA pipeline \citep{Casa_desc} to perform data reduction and again extracted the flux density after 1-2 rounds of self-calibration and imaging with \textit{tclean} using \textit{imfit}. The final epoch at 1071 days was taken at C configuration, where the beam size at S band (7x7 arcsec$^2$) and C band prevented the extraction of the SN flux density (the SN had grown too faint to be detectable above the noise from the host galaxy emission). We thus report 3$\sigma$ upper limits with the same methodology used for GMRT data at this last multi-frequency VLA epoch. We show radio images from the GMRT and VLA in  Figure \ref{fig:Overall_images}. The radio lightcurve of SN ~2019yvr at multiple frequencies is shown in Figure \ref{fig:Fitall}.

\subsection{\textit{Swift} and \textit{Chandra} X-ray Observations}
\begin{deluxetable*}{lcccccc}
\tablecaption{Summary of X-ray Observations of SN\,2019yvr \label{tab:xray_obs}}
\tablehead{
\colhead{Instrument} &
\colhead{Days post-explosion} &
\colhead{Count Rate (cts/s)} &
\colhead{Unabs. 0.2-10 keV Flux $(\rm erg\,s^{-1}\,cm^{-2})$} &
\colhead{Exp. Time (s)}
}
\startdata
Swift-XRT & $18.1 \pm 6.65$ & $<(3.34 \pm 0.52)\times10^{-3}$ & $<9.12 \times 10^{-14}$ &  12900 \\
Chandra ACIS-S & $42.7 \pm 0.1$ & $(1.15 \pm 0.36)\times10^{-3}$ & $(2.09 \pm 0.65)\times10^{-14}$ & 9920 \\
Swift-XRT & $68.9 \pm 9.44$ & $<(2.85 \pm 0.36)\times10^{-3}$ &$<1.33 \times 10^{-13}$ & 23470 \\
Swift-XRT & $196.4 \pm 14.4$ & $<(3.30 \pm 0.36)\times10^{-3}$ & $< 1.3 \times 10^{-13}$& 28200 \\
Swift-XRT & $391.2 \pm 21.9$ & $<(1.76 \pm 0.55)\times10^{-3}$& $< 8.2 \times 10^{-14}$  & 6253 \\
Swift-XRT & $1682.1 \pm 14.7$ & $<(2.40 \pm 1.14)\times10^{-3}$ & $< 1.12 \times 10^{-13}$ & 2341 \\
\enddata
\tablecomments{Days post-explosion taken as days relative to December 22, 2019 (MJD 58839) \citep{Ferrari_2024}. The count rate includes host galaxy emission for Swift-XRT observations, which was then subtracted using pre-explosion imaging for the supernova flux measurement.}
\end{deluxetable*}

SN 2019yvr was observed at X-ray wavelengths with the \textit{Swift-XRT}  at over 30 individual epochs up to 4 years post-explosion. It was also observed once with the \textit{Chandra} X-ray telescope at 42 days post-explosion (PI D. Pooley, ObsID 21307). The host galaxy of SN 2019yvr, NGC 4666, is a known X-ray bright AGN \citep{Persic_2004}, although its multi-wavelength properties are not well-studied. With the SN at $\sim$ 15 arcsec from the nucleus of the host galaxy, the galaxy contaminated the supernova flux in all \textit{Swift-XRT} observations. Fortunately, there are pre-explosion \textit{Swift-XRT} observations of the host galaxy from 2014--2016, which we combined to create a 100 ksec exposure (Obs ID 33545,9194,86019) which we used to attempt to remove host galaxy contamination. We do not use the \textit{Chandra} post-explosion epoch to constrain the host galaxy, given that the low exposure time did not allow for robust modeling of the AGN, opting instead to use the pre-explosion \textit{Swift} data to determine the host flux and constrain any post-explosion SN flux in \textit{Swift} data.

We reduced all data using \textit{xrtpipeline}. For the pre-explosion data, we find a count rate from the host galaxy of $2.9\times 10^{-3}$ counts/sec in a 20 arcsecond region centered on the SN (with the background defined with a 150 arcsecond source-free region).  We show the region around the SN in Figure \ref{fig:Overall_images}.  We fit a power law to the data with \textit{xspec} and find a photon index $\Gamma=1.53 \pm 0.1$. The initial best-fit column density is consistent with the reported value at the galaxy location (and thus we freeze this parameter in the final fit) of $1.75 \times 10^{20}\, \rm cm^{-2}$ \citep{Willingale_2013}. Using \textit{cflux}, we obtain an unabsorbed 0.2-10 keV host galaxy flux  of $1.02_{-0.15}^{+0.14}\times 10^{-13} \mathrm{erg\, \rm{cm^{-2}s^{-1}}}$. 

Since the \textit{Swift-XRT} individual exposures are of short durations, we proceeded to bin the post-explosion observations (OID 13038) into 5 different epochs, attempting to ensure $\delta t/t<0.2$ but also to accumulate enough counts in each binned exposure to allow for spectral fitting. Our first four epochs span from $18-391$ days post-explosion, with one later epoch at 1682 days (see Table \ref{tab:xray_obs}). We extract a spectrum from the same region as was used for the pre-explosion analysis. We applied the galaxy model from the pre-explosion data and attempted to add a SN power-law model, assuming the galaxy flux would be relatively unchanged. There is no statistical evidence for an additional flux component in any of the \textit{Swift-XRT} data from an F-test using the change in $\chi^2$. Using the XSPEC task \textit{steppar}, we obtain 1$\sigma$ limits (from finding where $\Delta \chi_{\nu}^2=1$) on the flux from the SN (obtaining 3$\sigma$ limits was not possible due to low counts). At the last two epochs, there were not enough counts to perform any 2-component fit. We thus use the host galaxy flux as a conservative upper limit for these last two epochs. The SN does not contribute any flux $ >10^{-13} \mathrm{erg\, \rm{cm^{-2}s^{-1}}}$ at any epoch. The flux limits are reported in Table \ref{tab:xray_obs}.  

As mentioned, SN 2019yvr was also observed by \textit{Chandra} at 42 days post-explosion (\dataset[DOI: 10.25574/cdc.640]{https://doi.org/10.25574/cdc.640}). With \textit{Chandra}'s superior resolution and sensitivity, the SN was isolated and detected with a 0.2--10 keV count rate of 1.15 $\times 10^{-3}$ counts/s. The spectrum had very low counts, and thus we used C-statistics to measure goodness of fit and used the \textit{error} command to find errors. Using \textit{xspec}, we fit the spectrum with a power law, with no difference in $\chi^2$ for any other model.  We assume the galactic column density along the line of sight, given there is no evidence for significant additional column density from fitting. We find a best-fit $\Gamma=1.98^{+1.81}_{-0.38}$. Using $\Gamma$ and the count rate, we use PIMMS \footnote{https://cxc.harvard.edu/toolkit/pimms.jsp} to find a flux of $(2.09 \pm 0.65) \times 10^{-14}\,\mathrm{erg\, \rm{cm^{-2}\,s^{-1}}}$. This flux gives an $L_{\rm X,0.3-10 keV}=(5.38\pm 1.99) \times 10^{38}$ erg\,s$^{-1}$. The measured SN luminosity and flux is reported in Table \ref{tab:xray_obs}.

\section{Data Analysis}\label{sec:Analysis}
We modeled the radio and X-ray data in a CSM interaction scenario to understand the mass-loss history of SN~2019yvr. We note that we add a 10$\%$ systematic error in quadrature to all radio flux  density errors \citep{BaerWay2025} for our modeling to account for underestimation of flux errors from \textit{imfit}.
\begin{figure*}
    \centering
    \includegraphics[width=18 cm, height= 6 cm]{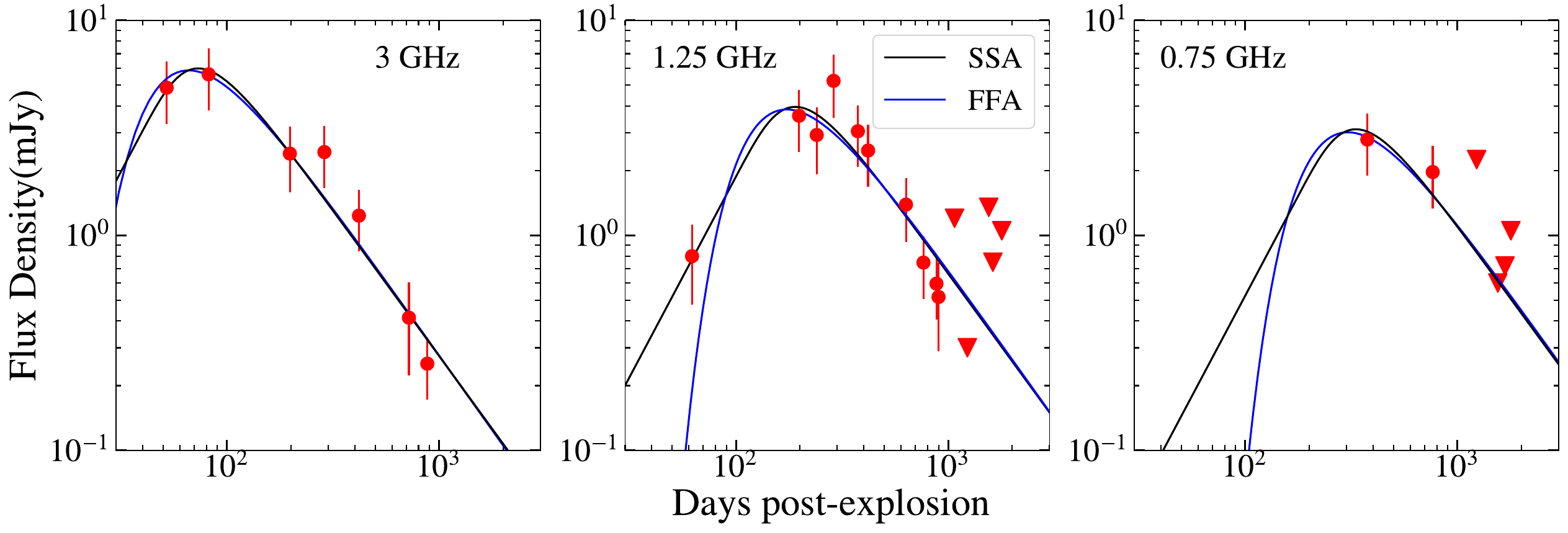}
    \caption{The radio lightcurves of SN 2019yvr at selected radio frequencies with the best-fitting SSA and FFA model to the entire dataset in frequency and time. We do not fit non-detections.}
    \label{fig:Fitall}
\end{figure*} 
\subsection{Radio analysis} 
Radio emission from ejecta-CSM interaction is expected to be synchrotron radiation from electrons in an amplified magnetic field, with the radiation attenuated either by the synchrotron electrons themselves in synchrotron self-absorption (SSA) or by external ionized CSM in free-free absorption (FFA). SSA and FFA can both be parametrized as detailed by e.g. \citet{Chevalier_1998,Weiler_1986} into forms describing the optically thick (flux rising) and optically thin (flux declining) evolution. For the self-similar formalism developed by \cite{Chevalier_1998}, one key assumption is that $\nu_{\rm m}<\nu_{\rm a}<\nu_{\rm c}$, where $\nu_{\rm m}$ is the minimum characteristic synchrotron frequency, $\nu_{\rm a}$ is the absorption frequency and $\nu_{\rm c}$ is the cooling frequency, with both inverse Compton cooling and synchrotron cooling possible. In order to calculate the relevant frequencies, we use the formulae from \cite{Nayana_2025}. We use a range of values measured in other stripped-envelope SNe for the radius and magnetic field of the shock $\rm R_{sh}$  and $\rm B_{sh}$ \citep{Bietenholz_2021} at each epoch, and estimates of the bolometric luminosity from \cite{Kilpatrick_2021}. We find that $\nu_{\rm m}<\nu_{\rm a}<\nu_{\rm c}$ at all epochs of observations for any reasonable $\rm B_{sh}/R_{sh}$ and any fraction of energy in relativistic electrons $\epsilon_{e}$. We thus proceed using the self-similar \cite{Chevalier_1998} formalism.

\subsubsection{Model Description}
For FFA, $F(\nu,t)$ can be parametrized as: \citep{Chevalier_1982,Chandra_2020, Weiler_1990}
\begin{equation}
F(\nu,t)_{\rm{FFA}} = K_{1} \left( \frac{\nu}{5 \hspace{0.1cm} \mathrm{GHz}} \right)^{-\alpha} 
\left( \frac{t}{\mathrm{1000 \hspace{0.1cm} days}} \right)^{-\beta} 
e^{-\tau_{\mathrm{FFA}}(\nu,t)} ,
\label{eq:FFA}
\end{equation}
Here, $K_{1}$ is the flux density normalization, $\alpha$ is the optically thin spectral index and $\beta$ is the temporal index of the flux density evolution. $\alpha$ is tied to the electron energy index $p$ by $p=2\alpha+1$ \citep{Chevalier_1998}. 
The FFA optical depth is given by \begin{equation}\tau_{\rm{FFA}}(\nu,t)=K_{2}\left(\frac{\nu}{5\hspace{0.1 cm} \mathrm{GHz}}\right)^{-2.1}\left(\frac{t}{1000 \hspace{0.1 cm}{\rm days}}\right)^{-\delta},\label{eq:FFA_tau}\end{equation} 
where $\delta$ describes the evolution of the optical depth, and
$K_{2}$ is the optical depth normalization. 

In the SSA model, the flux density can be written as \citep{Weiler_1986,Chevalier_1998} \begin{align}
F(\nu,t)_{\rm{SSA}} =\;& K_{1}\left(\frac{\nu}{5 \hspace{0.1 cm} \mathrm{GHz}}\right)^{2.5}
\left(\frac{t}{\rm{1000 \hspace{0.1 cm} days}}\right)^{-\beta'} \nonumber \\
&\times \left(1 - e^{-\rm{\tau_{SSA}(\nu,t)}}\right)
\label{eq:SSA}
\end{align}

The SSA optical depth is \begin{equation}\tau_{\rm{SSA}}=K_{2}\left(\frac{\nu}{\mathrm{5 \hspace{0.1 cm} GHz}}\right)^{(-\alpha-2.5)}\left(\frac{t}{\rm{1000 \hspace{0.1 cm} days}}\right)^{(-\beta'+\beta)}\label{eq:SSA_tau}\end{equation} 
$K_{2}$ again corresponds to the optical depth normalization and $\alpha$ is again the optically thin spectral index.
$\beta'$ and $\beta$ denote the temporal evolution of flux densities in the optically thick ($F \propto t^{\beta^{'}}$) and thin phase ($F \propto t^{\beta}$).
\subsubsection{Joint fit to all radio data}
We performed MCMC fits to the full radio dataset in frequency and time using equations \ref{eq:FFA} and \ref{eq:SSA} with all variables as free parameters, considering SSA-dominant and FFA-dominant cases. We used \textit{emcee} \citep{Foreman_2013} with broad priors on the various parameters, a 4000 step burn-in and 10000 total iterations with 200 walkers. We check for convergence using the stretch statistic. We use this setup for all model fitting in this work. The microscopic parameters describing the shock i.e., the electron energy density and magnetic field energy density, $\epsilon_{e}$ and $\epsilon_{B}$, remain fixed in these overall fits. We also note that by fitting to all epochs at once, we are implicitly assuming a CSM density profile that has a single power-law form (i.e. $\rm \rho_{CSM}\propto r^{-s}$) with no sharp changes.

\begin{figure}[htbp!]
    \centering
    \includegraphics[width=8.7 cm, height =11.5 cm]{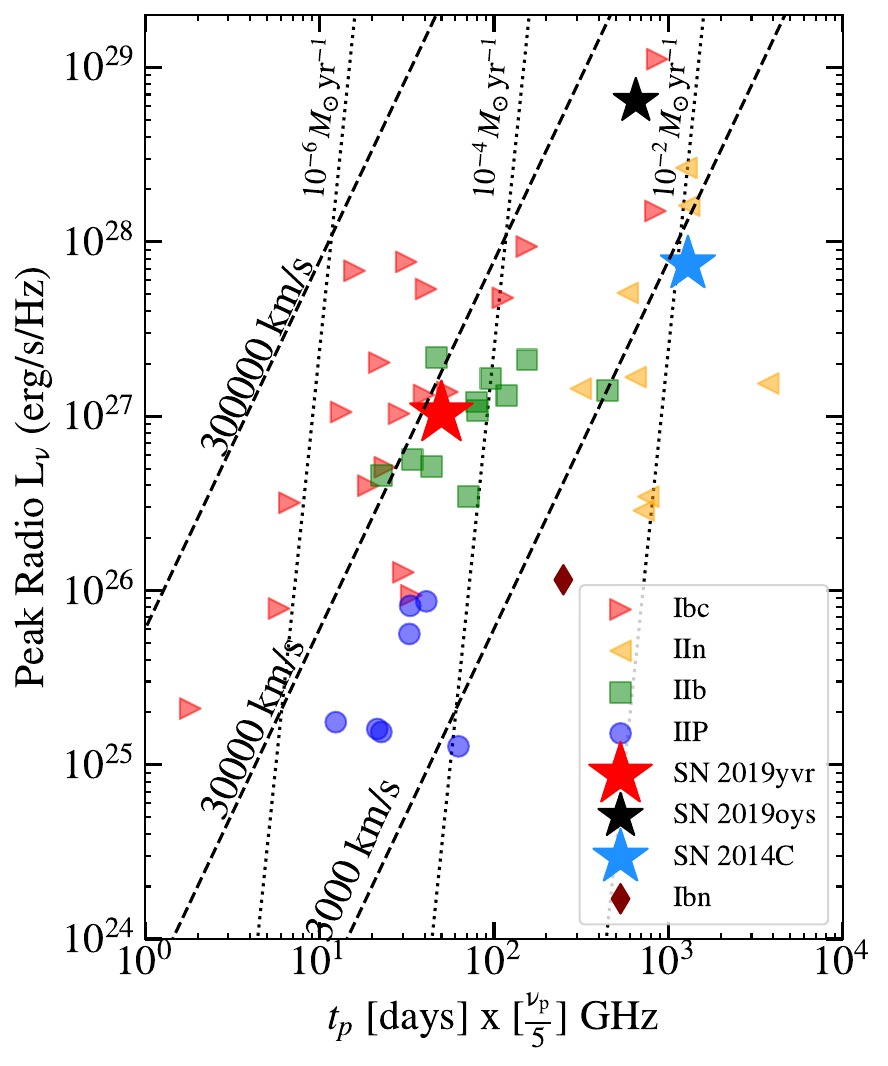}
    \caption{A Chevalier diagram (peak spectral radio luminosity $L_\nu$, vs peak time, $t_p$ normalized to 5 GHz; \cite{Chevalier_1998}) with a variety of SESNe (data taken from \cite{Bietenholz_2021}, \cite{Anderson_2017}, \cite{Sfaradi_2024} and \cite{RBW_2025}). The peak times are directly taken from collected data (i.e. not modeled). We plot other transition objects such as SN 2014C and SN 2019oys for comparison. For SN 2014C, we plot the secondary peak  (after it transitioned to a type IIn at optical wavelengths). The dashed and dotted lines represent lines of constant velocity and constant $\rm \dot{M}$ (for a 100 km/s CSM speed), respectively.}
    \label{fig:Chevalier}
\end{figure}

\begin{deluxetable*}{cccccccc}
\tablecaption{SSA vs. FFA fits details for the full radio dataset for SN\,2019yvr} \label{tab:params}
\tablehead{\colhead{Model}&
\colhead{K1} & \colhead{K2} & \colhead{$\alpha$} & \colhead{$\beta$} & \colhead{$\delta$} & \colhead{$\beta'$} & \colhead{$\chi_{\nu}^2$}
}
\startdata
FFA&$0.16_{-0.01}^{+0.01}$ &$0.001_{-0.0002}^{+0.0003}$&$	1.06_{-0.03}^{+0.04}$ &$1.39_{-0.04}^{+0.04}$ &	$1.89_{-0.09}^{+0.09}$& \nodata &$5.77$\\
SSA&$4130_{-1052}^{+1035}$&$0.00004_{-0.00001}^{+0.00001}$&$1.01_{-0.03}^{+0.03}$&$1.36_{-0.04}^{+0.04}$& \nodata & $1.84_{-0.13}^{+0.10}$ & $5.39$ \\
\enddata
\label{tab:Initial_fitting}
\tablecomments{The parameters for SSA and FFA are explained in section \ref{sec:Analysis}.}
\end{deluxetable*}
We find from our fits that a pure SSA model is slightly favored (see Table \ref{tab:Initial_fitting} for all best-fit parameters) with $\chi_{\nu}^2=5.39$, as compared to the FFA fit that gives $\chi_{\nu}^2=5.77$. A view of the lightcurves at low frequencies where the optically thick and optically thin phases are sampled is shown in Figure \ref{fig:Fitall}. The SSA model appears to be the slightly better fit (see especially the 1.25 GHz fit). Furthermore, when plotting the 1.25 GHz peak in a Chevalier diagram of peak luminosity vs peak time \citep{Chevalier_1998}, the implied shock speed (which can be inferred from this combination of peak radio luminosity and time to peak) is $\sim$ 30000 km/s-suggesting a high shock-speed SSA scenario (where the ejecta are not decelerated to very low velocities as would be the case if there was significant FFA  -- see Figure \ref{fig:Chevalier}). We thus proceed with a pure SSA model as the best-fitting model as our dataset does not reveal strong evidence for FFA.

However, the overall fits in frequency and time do not suggest an entirely homogeneous CSM based on the $\chi^2$ and the over/underprediction of some of the 1.25 GHz points. We thus opt to fit an SSA model at individual epochs to consider sudden changes in the circumstellar environment across epochs and understand the evolution of the shock parameters in detail. Our data at individual epochs do not always reveal the turnover from optically thick to optically thin emission. While the GMRT data was always taken within $\sim$ 30 days of the VLA data, it is not contemporaneous, especially at early times. We can use the overall SSA model to extrapolate band 4 GMRT (0.75 GHz) datapoints to the exact epoch of VLA observations from 0-500 days, inserting $\nu$ and using all best-fit parameters from Table \ref{tab:Initial_fitting} in Equation \ref{eq:SSA}. After $\sim$ 500 days, the band 4 points themselves are optically thin as well. We thus can only construct four epochs for modeling which contain optically thick and thin datapoints (see Figure \ref{fig:SSA}): One at 52 days (where we also extrapolate C and X band datapoints given the S band detection), one at 198 days, one at 287 days, and one at 419 days.

\subsubsection{Fits to individual epochs}
\begin{figure*}[htbp!]
    \centering
    \includegraphics[width=18 cm, height =5 cm]{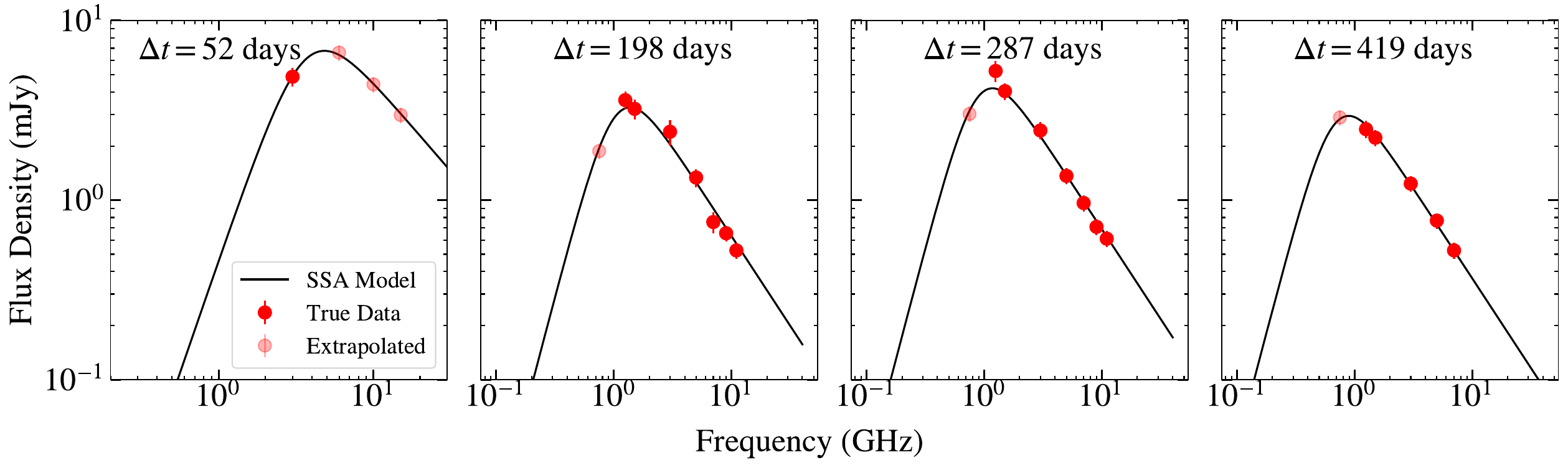}
    \caption{SSA single-epoch broken power-law fits for the four epochs we fit. We extrapolate points from the near-contemporaneous GMRT low-frequency data points at each epoch. For the exact details of the extrapolation calculation, see section \ref{sec:Analysis}.}
    \label{fig:SSA}
\end{figure*}

SSA can be parametrized at a single epoch by a broken power law: 
\begin{equation}
F_{\nu}=F_{\rm brk}[(\frac{\rm\nu}{\rm \nu_{brk}})^{\alpha_{1}/s}+(\frac{\rm \nu}{\rm \nu_{brk}})^{\alpha_{2}/s}]^{S}
\label{eq:SSA_BPL}
\end{equation}
\citep{Demarchi_2022} where $F_{brk}$ is the intersection between the asymptotic power-laws representing optically thin and thick portions of the SED \citep{Demarchi_2022}, $\alpha_{1}$ and $\alpha_{2}$ represent the optically thin and thick spectral indices, $S$ is the smoothing parameter which reflects how abruptly the SED turns over from optically thick to optically thin and $\rm \nu_{brk}$ is the peak frequency where $\tau_{\rm SSA}=1$. In general, $\alpha_{2}=2.5$ for SSA \citep{Chevalier_1998} and $\alpha_{1}=-\frac{(p-1)}{2}$. Having already constrained $\alpha_{1}\sim -1$ from the overall SSA fit (see Table \ref{tab:Initial_fitting}), we opt to fix $\alpha_{2}=2.5$ as should always be the case for SSA \citep{Chevalier_1998}. We fit with both $S=-1$ fixed (as is expected in a smooth SSA spectrum) and with $S$ varying \citep{Demarchi_2022}. From an F-test across the four individual epochs we identified, we find no statistical evidence for $S \neq -1$. We thus fit the four radio SEDs with equation \ref{eq:SSA_BPL} with $S$, $\alpha_{1}$ and $\alpha_{2}$ fixed to obtain $F_{\rm brk}$ and $\nu_{\rm brk}$ \citep{Demarchi_2022}.  The best-fit $\nu_{\rm \rm brk}$ and $F_{\rm brk}$ are given in Table \ref{tab:SSAindepoch}. We show the fits in Figure \ref{fig:SSA}, with $\chi^2_{\nu} \sim $1 at each epoch, suggesting a good fit.
We can then find shock properties such as the magnetic field and radius, as these properties directly set the frequency at which the peak flux occurs in an SSA spectrum and the peak flux itself \citep{Chevalier_1998}, using the best-fit $\rm F_{brk}$ and $\rm \nu_{brk}$ assuming $p=3$ \citep{Chevalier_1998}:
\begin{equation}
    \begin{split}
    R=8.8 \times 10^{15}f_{eB}^{-1/19}\left( \frac{f}{0.5}\right) ^{-1/19}\left(\frac{F_{brk}}{\mathrm{Jy}}\right)^{9/19}\times
    \\ \left(\frac{D}{\mathrm{Mpc}}\right)^{18/19}\left (\frac{\nu_{brk}}{5 \hspace{0.1 cm}\mathrm{GHz}}\right)^{-1} \mathrm{cm}
    \end{split}
    \label{R_eq}
\end{equation}
\begin{equation}
    \begin{split}
    B=0.58 f_{eB}^{-4/19}\left(\frac{f}{0.5}\right)^{-4/19}\left (\frac{F_{brk}}{\mathrm{Jy}}\right)^{-2/19}\times\\ \left(\frac{D}{\mathrm{Mpc}}\right)^{-4/19}\left(\frac{\nu_{brk}}{5 \hspace{0.1 cm} \mathrm{GHz}}\right) \mathrm{G}
    \end{split}
    \label{B_eq}
\end{equation}

\noindent
where $f_{\rm eB}=\frac{\epsilon_{e}}{\epsilon_{B}}$ denotes the ratio between the energy density of relativistic electrons ($\epsilon_{e}$) and the magnetic field energy density ($\epsilon_{B}$), and $f$ is the volume filling factor (which we assume to be 0.5 as is standard for radio SNe \citep{Weiler_1986}). We assume $f_{eB}$=10 as has been inferred in many SESNe \citep{Sfaradi_2025}, with further motivation for this choice given later in this section. Using these parameters and equations \ref{R_eq} and \ref{B_eq}, we calculate $ R$ and $ B$. We can then use these parameters to find the unshocked CSM density $\rm \rho_{CSM}$ at each epoch by asserting that some fraction $\epsilon_{B}$ of the total thermal energy density in the CSM $U_{th}=\frac{9}{8}\rho_{CSM}{\rm V}_{\rm sh}^2$ (where ${\rm V}_{\rm sh}$ is the shock velocity defined as $R/t$) will go into the magnetic field energy density ($\rm B^2/8\pi$):
\begin{equation}
    {\rm \frac{B^2}{9\pi}=\epsilon_{B}\rho_{CSM}V_{sh}^2}
    \label{eq:rho}
\end{equation}
\citep{Demarchi_2022}.
\subsubsection{Derived CSM and shock parameters}
\begin{figure}
    \centering
    \includegraphics[width=8 cm, height=5 cm]{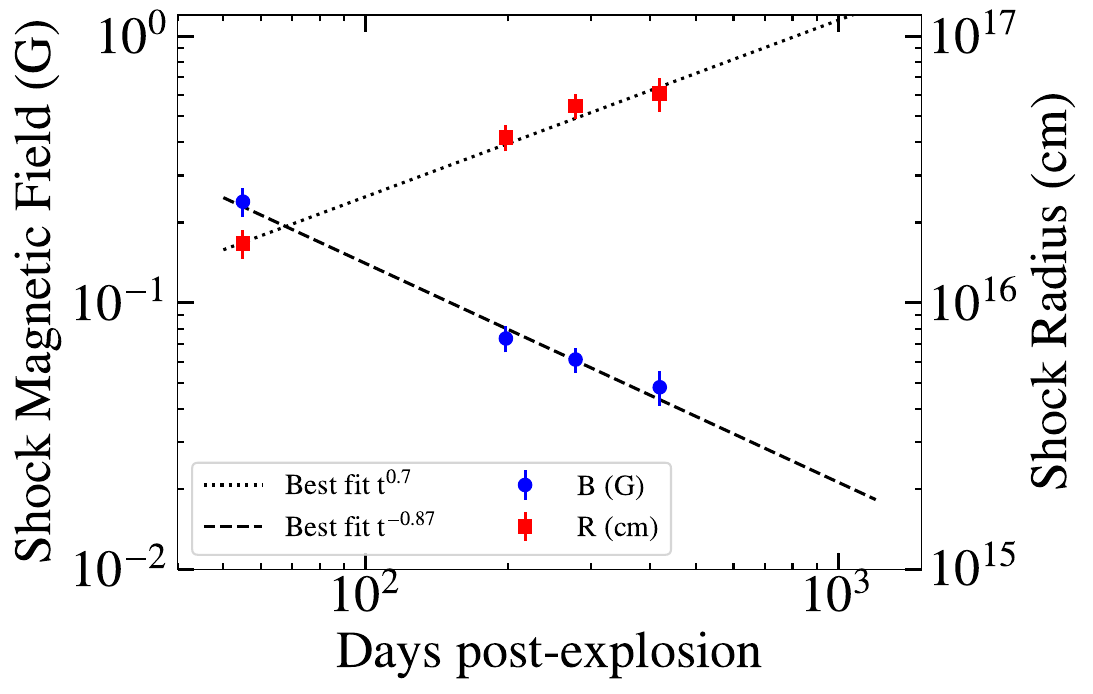}
    \caption{The evolution of the post-shock magnetic field and the shock radius from the best-fit single epoch SSA modeling described in Section \ref{sec:Analysis}. We show the best-fit power laws.}
    \label{fig:BR_plot}
\end{figure}

The shock radius, velocity (radius at each epoch divided by days post-explosion), and magnetic field values are given in Table \ref{tab:SSAindepoch}. The radius and magnetic field evolution are plotted in Figure \ref{fig:BR_plot}. The CSM radius increases from $(1.6-6) \times 10^{16}$ cm, while the magnetic field decreases from 0.24 to 0.05 Gauss over the first $\sim$ 400 days post-explosion. The shock is significantly decelerated by the CSM throughout the evolution as the velocity drops from $\sim$ 35000 km/s at 52 days to 17000 km/s at 419 days. Using the derived radii, we can obtain the deceleration parameter of the shock $m$ as the CSM decelerates the shock and $\rm R_{sh}$ goes as $t^{m}$. We performed an MCMC fit to the evolution of the radius and find $m=0.70_{-0.07}^{+0.13}$. We show the fit in Figure \ref{fig:BR_plot}. We also find that $B\propto t^{-0.87_{-0.18}^{+0.09}}$, and define this temporal exponent of the B-field as $\alpha_{B}$. 

We can use the magnetic field and shock velocity values derived at each epoch from equations \ref{R_eq}  and \ref{B_eq} to calculate $\rm \rho_{CSM}$ values assuming some $\rm \epsilon_{B}$ (see equation \ref{eq:rho}). We calculated $\rm \rho_{CSM}$ in two scenarios: one with $\epsilon_{B}=0.01$ and thus $f_{eB}=10$ as has been seen in many SESNe \citep{Sfaradi_2025}, and another with $\epsilon_{B}=\epsilon_{e}=0.1$ (equipartition). We find that assuming $\epsilon_{B}=0.01$ and $f_{eB}=10$ so $\epsilon_{e}=0.1$ gives rough consistency with measurements from the X-ray detection (see Section \ref{sec:Disc}) and the CSM inferences at optical wavelengths \citep{Ferrari_2024}, and thus we adopt this assumption. The $\rm \rho_{CSM}$ values range from $10^{-20}-10^{-21}\rm  g/cm^3$ and are given in Table \ref{tab:SSAindepoch}. These $\rm \rho_{CSM}$ values can then be used to calculate the power-law density exponent of the CSM, which we define as $s$ (the \cite{Chevalier_1998} self-similar formalism we use implicitly assumes $\rm \rho_{CSM} \propto r^{-s}$). From fitting to the $\rm \rho_{CSM}$ values as a function of radius, we find $s=1.54 \pm 0.31$.

The $s$ and $m$ values can be related to the density profile of the outer ejecta $\rho_{ejecta}\propto r^{-n} $ assuming self-similarity by $\rm m=\frac{n-3}{n-s}$ \citep{Chevalier_17}. Our derived $s$ and $m$ values imply an outer ejecta density exponent of $n_{outer} \sim 7$ in this formulation.  This $n$ value is slightly shallow compared to expectations for typical SESNe \citep{Chevalier_17}, and the $m$ value implies a strongly decelerating shock. The derived CSM density profile exponent is \textbf{$s=1.54 \pm 0.31$} (not wind-like, which would have given $s=2$) and implies an excess of material at larger radii compared to CSM formed from constant mass loss.

With $\rm \rho_{CSM}$ and $s$, we then calculate the mass-loss rate that produced the CSM generating the shock at each epoch $\rm \dot{M}$ using the following formula:
\begin{equation}
    \rm \rho_{CSM}=\frac{\dot{M}}{4\pi r_{0}^{2}v_{CSM}}\left( \frac{r_{0}}{r}\right)^{s}
    \label{eq:Mdot_from_rho}
\end{equation}
\citep{Frannson_96}
where $r_{0}$ is a reference radius at which $\frac{\dot{M}}{v_{CSM}}$ can be assumed. We choose the radius derived at the first radio epoch as the reference radius. The CSM speed is also needed to calculate $\rm \dot{M}$ from $\rho_{CSM}$. The CSM speed for SN 2019yvr is not constrained by optical spectra; only an intermediate-width shock-driven H$\alpha$ line of FWHM $\sim$ 2000 km/s was seen post-transition \citep{Ferrari_2024}. We assume a 100 km/s CSM speed, as was also assumed by \cite{Kilpatrick_2021}, given that the progenitor is likely in a binary system (with the binary system driving the mass transfer based on the analysis of \cite{Sun_2021}) and the fact that binary systems with stable mass transfer are expected to have outflow speeds $\sim$ 100 km~s$^{-1}$ \citep{Sun_2021,Mandal_2026}. We calculate mass-loss rates from $\sim 1-4 $ $\times 10^{-5} \mathrm{M_{\odot}yr^{-1}}(\frac{\rm v_{CSM}}{\rm 100 km s^{-1}})$ across our four epochs (see Table \ref{tab:SSAindepoch}).  

We further find that the radio flux densities measured at 880 and 1777 days are consistent with the extrapolation of the overall SSA model (with parameters detailed in Table \ref{tab:Initial_fitting}). We can thus use the best-fit $m$ and $\alpha_{B}$ to find shock parameters at later times. Extrapolating B, R and $V_{sh}$ out to these late epochs (using the parameters from the epoch at 419 days as a reference point) and using equations \ref{eq:rho} and \ref{eq:Mdot_from_rho}, we find mass-loss rates of $(3-6) \times 10^{-5} \mathrm{M_{\odot}yr^{-1}}$ that are similar to the values derived at earlier epochs via SED fits. The detection at 1722 days would also imply that the CSM extends to at least $1.6\times 10^{17}$ cm based on the derived $m$, provided that the extrapolation for the evolution of the radius holds.

As an independent check on the mass-loss history, we can also calculate what the implied total mass of the CSM would be given the measured deceleration of the shock. The deceleration of the shock from 35000 to 17000 km~s$^{-1}$ in the first four epochs for an ejecta mass of 2 $\rm M_{\odot}$ \citep{Ferrari_2024} would require a CSM mass of 2.2 $\rm M_{\odot}$ \citep{Nayana_2023ixf} if the ejecta are almost entirely swept up at these epochs, which is not consistent with the integrated mass-loss rate over these first four epochs. We suggest that the slightly increasing mass-loss rate (and hence non-$r^{-2}$ CSM density profile) at earlier phases pre-explosion causes the strong deceleration.


\begin{deluxetable*}{cccccccc}
\tablenum{3}
\tablecaption{SSA single-epoch Broken Power-Law fits to individual epochs for SN\,2019yvr \label{tab:SSAindepoch}}
\tablehead{
\colhead{Epoch} & \colhead{$F_{\rm brk}$ (mJy)} & \colhead{$\nu_{\rm brk}$ (GHz)} & \colhead{$\rm R_{sh}$ (cm)} & \colhead{$\mathrm{\dot{M}\,(M_{\odot}yr^{-1}})$} & \colhead{$\rm V_{\rm sh}$ (km/s)} & \colhead{B (G)} & \colhead{$\rho_{\rm CSM}$ (g cm$^{-3}$)}
}
\startdata
52  & $12.30_{-0.70}^{+0.70}$ & $3.72_{-0.21}^{+0.23}$
& $(1.7 \pm 0.2) \times 10^{16}$ & $(0.9 \pm 0.1) \times 10^{-5}$ & $35000 \pm 4000$ & $0.24 \pm 0.03$ & $(1.6 \pm 0.2) \times 10^{-20}$ \\
198 & $5.95_{-0.31}^{+0.31}$ & $1.06_{-0.05}^{+0.05}$
& $(4.2 \pm 0.5) \times 10^{16}$ & $(1.7 \pm 0.2) \times 10^{-5}$ & $24000 \pm 3000$ & $0.07 \pm 0.01$ & $(3.2 \pm 0.4) \times 10^{-21}$ \\
287 & $7.61_{-0.36}^{+0.35}$ & $0.91_{-0.04}^{+0.04}$
& $(5.5 \pm 0.6) \times 10^{16}$ & $(2.6 \pm 0.3) \times 10^{-5}$ & $23000 \pm 2000$ & $0.06 \pm 0.01$ & $(2.6 \pm 0.3) \times 10^{-21}$ \\
419 & $5.38_{-0.37}^{+0.42}$ & $0.68_{-0.06}^{+0.06}$
& $(6.1 \pm 0.9) \times 10^{16}$ & $(3.9 \pm 0.6) \times 10^{-5}$ & $17000 \pm 2000$ & $0.05 \pm 0.01$ & $(2.9 \pm 0.5) \times 10^{-21}$ \\
\enddata
\tablecomments{We take $\alpha_{1}=-1,\, \alpha_{2}=2.5$, $S=-1$, and $\epsilon_{B}=0.01,\, \epsilon_{e}=0.1$ to derive $\rm \dot M$, 
$\rm V_{sh}$, R and B values. We also assume a CSM velocity of 100 km/s for the $\rm \dot{M}$ column. The errors given are purely statistical, not accounting for systematic errors arising from assumptions about $\rm v_{CSM}$ or $\rm \epsilon_{B}$. Epoch is defined as days post-explosion, taken relative to the December 22, 2019 date estimated by \cite{Ferrari_2024}.}
\end{deluxetable*}
\subsection{X-Ray analysis}
The X-ray detection at 42 days allows for another constraint on the mass-loss rate, despite the fact that the origin of the emission (i.e. non-thermal or thermal) is not easily constrained by the X-ray data alone with limited counts. Given the radio modeling, we can attempt to constrain whether the X-ray emission is due to inverse Compton X-ray emission, synchrotron emission or thermal free-free emission, with all three cases plausible for SESNe \citep{Dwarkadas_2025}. Using the extrapolated SSA synchrotron flux at 42 days, we immediately rule out synchrotron X-ray emission as the predicted 1 keV flux would be two orders of magnitude lower than the observed flux (and a cooling break between 15 GHz and 1 keV is quite likely, which would only exacerbate this difference). 

Furthermore, using the formalism of \cite{Boaz_2012} we use the radio SSA model+X-ray detection to calculate the expected radius and magnetic field in the shock if the X-ray emission was due to inverse Compton scattering. Using the bolometric luminosity estimates from \cite{Sun_2021} and our own X-ray and radio measurements, we find that $R$ and $B$ are over- or underpredicted (compared to the best-fit single-epoch day 52 values found in Table \ref{tab:SSAindepoch}) by a factor of 2--3, suggesting that an inverse Compton mechanism alone cannot explain the X-ray emission fully. In addition, strong inverse Compton emission should cause strong cooling at radio frequencies \citep{Maeda_2012}, which is not seen in the data. We thus conclude that the X-ray emission is likely thermal.

Having determined that the X-ray emission is thermal, we attempt to constrain whether the X-ray-emitting shocks are radiative or adiabatic. We use equations 14 and 18 of \citet{Chevalier_17} and the radio-derived mass-loss rate to find that the reverse and forward shocks are adiabatic past the first day post-explosion (regardless of CSM or ejecta density profile), and thus we can assume we are seeing emission from an adiabatic reverse shock (we would only see a strong adiabatic forward shock contribution if the reverse shock was radiative \citep{Frannson_96}). If we assume the temperature of the X-ray emitting plasma T=5 keV, we can use the following equation from \citet{Frannson_96}: 
\begin{equation}
\begin{split}
L_{\rm rev}(1\,\text{keV}) 
&= 1.74 \times 10^{37}\,\xi_4
   \frac{(n-3)(n-4)^2}{(3-s)^2(4-s)^2}
   T_{8}^{-0.24} \\
&\quad  e^{-0.0116/T_{8}} \xi\, C_{*}^{2} V_{4}^{3-2s}\,
\times \left( \frac{t_{d}}{11.57} \right)^{3-2s}
\end{split}
\label{eq:Mdot}
\end{equation}
where $C_{*}=\frac{\dot{M_{-5}}}{v_{w,10}}$ is the mass-loss rate over the wind velocity normalized to a 10 km/s wind and $\mathrm{10^{-5} M_{\odot}yr^{-1}}$ mass-loss rate, $t_{d}$ is the time since explosion and $T_{8}$ is the temperature of the shock in units of $10^{8}$ K. We find the spectral luminosity at 1 keV using the measured 0.3-10 keV luminosity from Table \ref{tab:xray_obs} (using PIMMS). We also assume $n=7;s=1.54$ from our radio results. Using the 1 keV spectral luminosity and a 100 km/s CSM speed, we find $\dot{M} =1.0 \pm 0.5 \times 10^{-4} \mathrm{M_{\odot}}\rm yr^{-1} \frac{ \rm v_{CSM}}{100~\rm km~s^{-1}}$ (assuming a $\sim$ 35000 km/s shock speed from the radio fit 10 days later). This $\rm \dot{M}$ is within 2$\sigma$ of all radio estimates. The X-ray upper limits from \textit{Swift} give conservative limits on the CSM density using the same methodology (implying only that the mass-loss rate $< 10^{-3}  \mathrm{M_{\odot}}\rm yr^{-1}$), and are consistent with the late-time radio results.

\section{Discussion}
\label{sec:Disc}

Based on our radio and X-ray data analysis and modeling, we established that SN 2019yvr was interacting with CSM created by mass loss at $\rm \sim10^{-5}-10^{-4} M_{\odot}yr^{-1}$ for a CSM speed of 100 km/s. Furthermore, the derived shock speed from the radio SED at day 52 ($\sim 0.1c$) suggests a compact progenitor star, as compact stars facilitate extreme acceleration in their outer layers \citep{Chevalier_2010}-but see SN 2011dh \citep{Soderberg_2012,Vandyk}. While there is no evidence for a significant change in CSM density corresponding to the onset of optical H$\alpha$ emission at 80-100 days post-explosion, there is evidence for an increase in progenitor mass-loss rate by a factor of $\sim$ 2-5 as the ejecta encountered more distant material that was lost longer before the explosion. The values of the mass-loss rate found in general are consistent with strong winds from a Wolf-Rayet star (which typically have higher wind speeds of $\sim$ 1000 km/s than we have assumed) or binary mass transfer at $\sim 10^{-5} \mathrm{M_{\odot}yr^{-1}}$  \citep{Grafener_2017,Ercolino_2025,Tsai_2026}. 
\subsection{SN 2019yvr in context with type-changing Supernovae }
\begin{figure}
    \centering
    \includegraphics[width=1.0\linewidth]{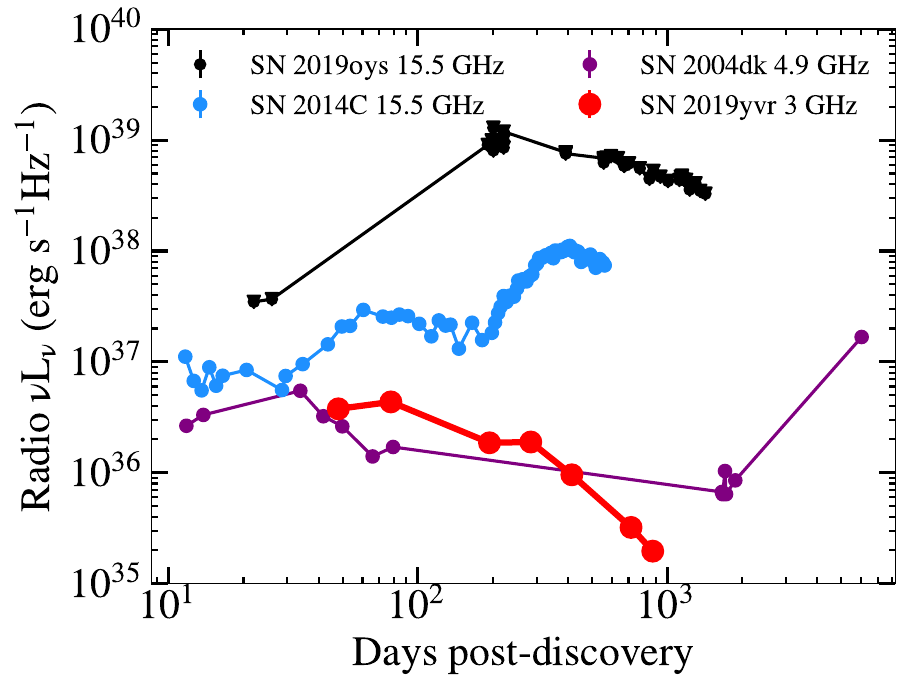}
    \caption{The radio lightcurve of SN 2019yvr at 3 GHz in comparison to the lightcurves at different frequencies for SNe 2004dk, 2014C and 2019oys. Data for SN 2019oys are from \cite{Sfaradi_2024} (with two additional datapoints from the VLA All-Sky Survey (VLASS) epoch 3 (extracting fluxes from the reduced image) and VLA proposal 24B-448 (reducing this data ourselves)). The data for SN 2014C are from \cite{Anderson_2017}. Data for SN 2004dk are from \cite{Wellons_2012,Balsubramanian_2021} with one additional point from VLA proposal 20B-279 (PI M.Stroh).}
    \label{fig:Radio_LCs}
\end{figure}

\begin{figure}
    \centering
    \includegraphics[width=8.5 cm, height = 6.5 cm]{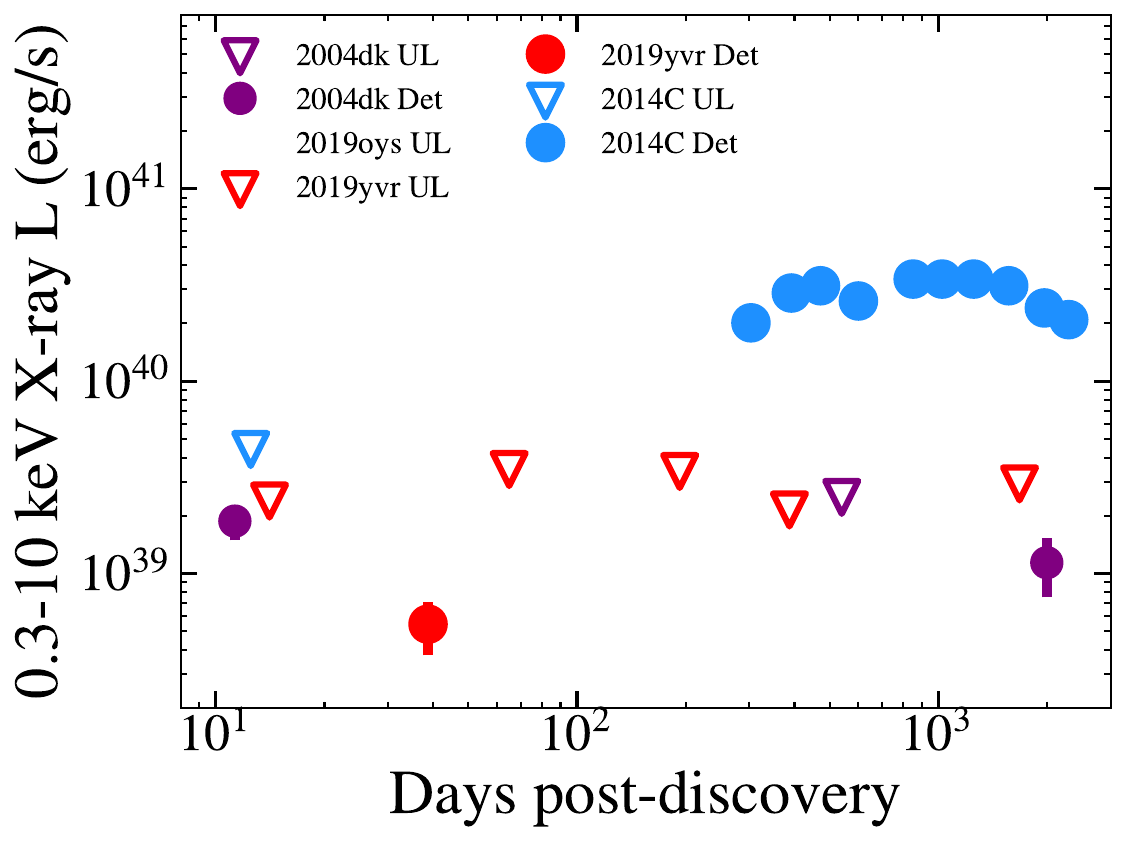}
    \caption{The X-ray evolution of SN 2019yvr from \textit{Swift} and \textit{Chandra} in context with SN 2014C, 2004dk and 2019oys (other SNe which underwent a $\rm Ib \rightarrow IIn$ transition). We show limits we derived from reanalysis of \textit{Swift} observations of SN~2019oys. We also show X-ray data points for SN 2004dk (rescaled to 0.3-10 keV luminosity using \texttt{PIMMS}) and for SN 2014C from \cite{Pooley_2019} and \cite{Brethauer_2022}, respectively.}
    \label{fig:Xrays}
\end{figure}

SN~2019yvr is one of four SNe (excluding SN 2001em due to ambiguity around the SN type at explosion -- see \cite{2001em_Alex,Chandra_2020}) that have definitively shown a transition from helium-rich early optical spectra to hydrogen-rich late-time spectra, presumably from running into hydrogen-rich CSM. The other three objects are SN 2004dk \citep{Wellons_2012,Mauerhan_2018}, SN~2019oys \citep{Sollerman_2020,Sfaradi_2024} and SN~2014C \citep{Danny_2014C,Margutti_2017,Thomas_2022,Brethauer_2022}. We show the radio and X-ray lightcurves of these three transition objects along with the lightcurves of SN 2019yvr calculated in this work in Figures  \ref{fig:Radio_LCs} and \ref{fig:Xrays}. While SNe 2004dk, 2019oys and 2014C have been studied extensively at radio and optical wavelengths \citep{Danny_2014C,Sfaradi_2024} and SNe 2004dk and 2014C have been studied at X-ray wavelengths \citep{Pooley_2019,Thomas_2022,Brethauer_2022}, there are only limited \textit{Swift} X-ray observations of SN~2019oys.  

We reduced and analyzed these X-ray observations of SN~2019oys. Using the same data reduction techniques as for SN~2019yvr, we binned the \textit{XRT} dataset collected on SN~2019oys into two separate epochs at 202 $\pm$ 5 (OID 13283) and 1024 $\pm$ 50 days (OID 96832) post-discovery. We obtain count rates at the SN location using \textit{ximage} (cross-checking with the \textit{Swift} online analysis tools \cite{Evans_2009}) of $(4.48 \pm 0.6) \times 10^{-3}$ counts/s and $(5.56\pm 1.5) \times 10^{-3}$ counts/s, respectively (SNR$>$ 3 at both epochs).  We note that these results differ from previous analysis done by \cite{Sollerman_2020}, who measured a higher count rate $\sim 17 \times 10^{-3}$ counts/s from the early epoch. The reason for the discrepancy is not clear.

SN~2019oys is not distant enough from its host at $\sim$ 6 arcsec separation (with \textit{Swift}'s PSF FWHM $\sim$ 6 arcsec) to be definitively uncontaminated, and there are no pre-explosion X-ray images to constrain the host flux. We note, however, that in the radio images of SN 2019oys \citep{Sfaradi_2024} (including L band images from VLA PID 24B-448) the 1.4 GHz radio luminosity from a 3$''$ region surrounding the host is $<10^{27}$ ergs/s/Hz (3$\sigma$ upper limit). This suggests it is highly unlikely that the host galaxy contains an AGN with $L_{X}>10^{40}$ ergs /s (see Figure 3 of \cite{Pennock_2025}). Given the count rates we measured, and assuming a power law as \cite{Sollerman_2020} did with $\Gamma=2$ and $N_{H}=9\times 10^{20} \rm cm^{-2}$, we obtain luminosities from the source near the SN given the 73 Mpc distance \citep{Sollerman_2020} of $> 10^{41}$ ergs/s. While we cannot definitively rule out galaxy emission, we suggest it is highly likely that this emission is produced in part or entirely by the SN ejecta-CSM interaction. We add these points as conservative upper limits in Figure \ref{fig:Xrays}, and note that this emission warrants further X-ray follow-up at very late times with \textit{Swift} or \textit{Chandra}. 


For SN~2004dk, we use the data already analyzed in the literature  \citep{Wellons_2012,Pooley_2019,Mauerhan_2018} and reduce one additional radio datapoint with the same methodology described for SN~2019yvr from proposal 20B-279 for comparison (PI M. Stroh). After showing early radio emission for the first $\sim$ 100 days consistent with lower-density CSM \citep{Wellons_2012}, SN 2004dk showed a late-time radio rebrightening (at $>$ 1500 days and continuing to $\sim$ 6000 days) at epochs past the current age of SN~2019yvr (see Figure \ref{fig:Radio_LCs}). 

\begin{figure}
    \centering
    \includegraphics[width=8 cm, height =6 cm]{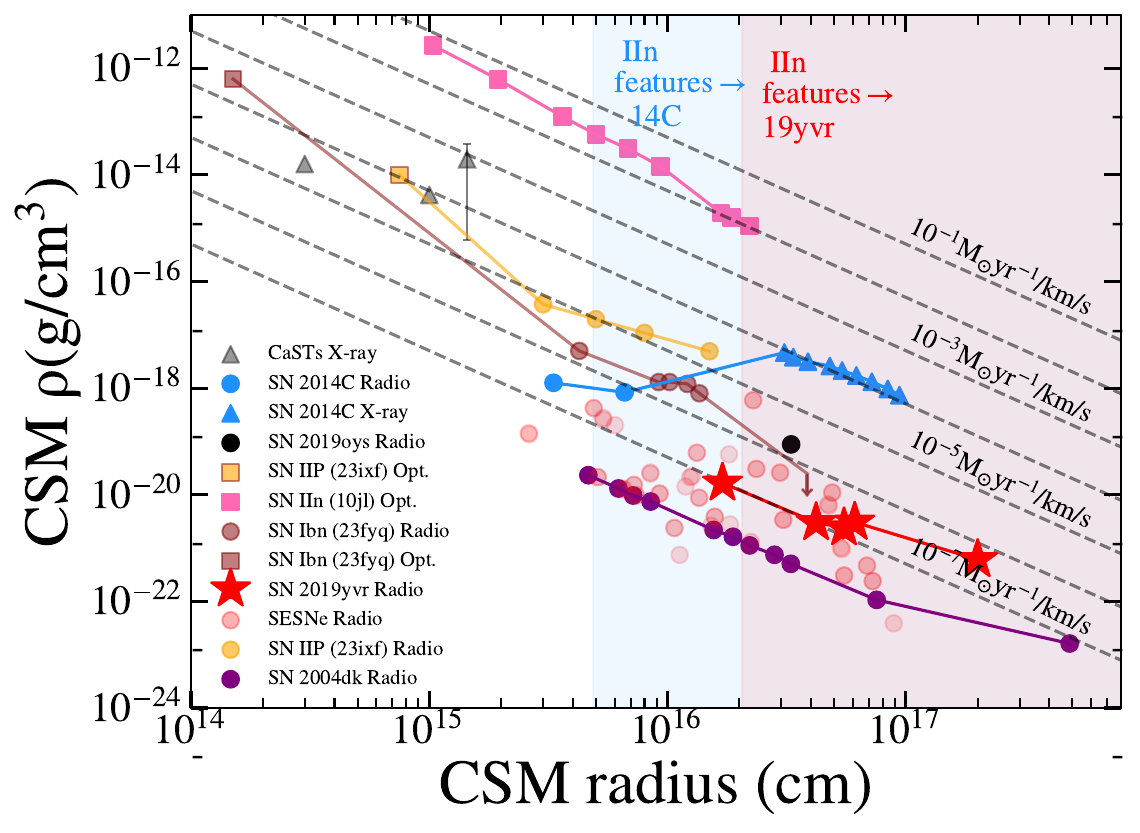}
    \caption{The CSM densities of SN~2019yvr inferred from our radio modeling in the context of other transition objects as well as other interacting and non-interacting SNe. Figure adapted from \cite{Kumar_2026}, see references therein for non-Ib $\rightarrow$ IIn objects. Data for SN 2014C are from \cite{Brethauer_2022} and \cite{Anderson_2017}; SN 2004dk from \cite{Wellons_2012}; and SN 2019oys from \cite{Sfaradi_2024}. Density profiles for constant $\rm \dot{M}/v_{CSM}$, i.e. a $\rho \propto r^{-2}$ profile, are plotted as diagonal lines. Shaded regions are added to note the CSM radii at which shock-driven H$\alpha$ features were seen in SNe 2019yvr/2014C \citep{Anderson_2017}.}
    \label{fig:CSM_rhocomp}
\end{figure}

\begin{figure}
    \centering
    \includegraphics[width=8 cm, height = 6 cm]{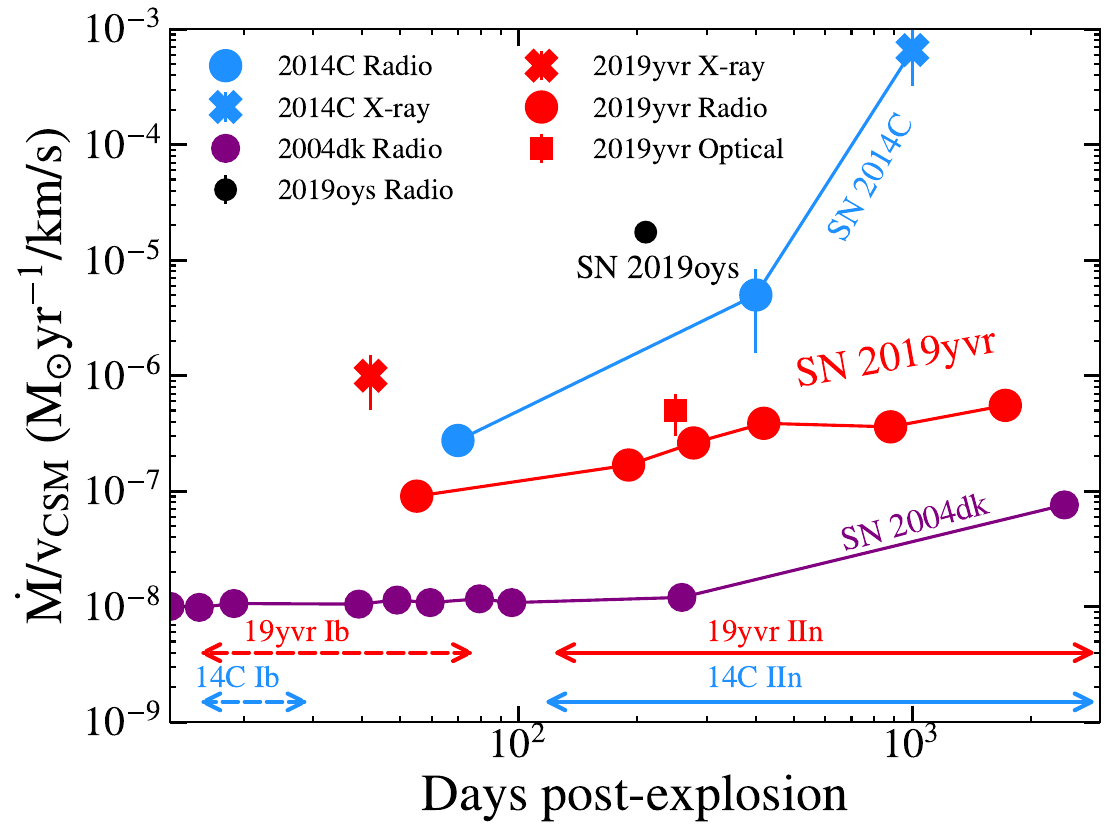}
    \caption{The $\rm \dot{M}/v_{CSM}$ inferred over time for SN~2019yvr measured from our radio SSA fits, optical modeling \citep{Ferrari_2024} and our X-ray modeling. The measured mass-loss rate at radio wavelengths increases slightly over time, but is not nearly as drastic as what was seen in SN~2014C. SN 2014C data are from SSA and FFA fits for the first two points \citep{Anderson_2017}, and we show the continuation with the blue line to a late point derived from X-ray data from \cite{Brethauer_2022}. We plot double-sided arrows to denote the timeframes over which SNe 2019yvr and 2014C had SN Ib-like and SN IIn-like features at optical wavelengths (see Table \ref{tab:Comp_4SNe} for details). Data for SN 2004dk are from \citet{Wellons_2012}. We also add the point from radio modeling of SN 2019oys \citep{Sfaradi_2024}.}
    \label{fig:ML}
\end{figure}

We plot the mass-loss histories (through the ratio of mass-loss rate/CSM velocity) and CSM densities derived for SN 2019yvr (from our work) and the other three $\rm Ib \rightarrow IIn$ objects (from \cite{Wellons_2012,Brethauer_2022,Sfaradi_2024}) in Figures \ref{fig:CSM_rhocomp} and \ref{fig:ML}. In Figure \ref{fig:CSM_rhocomp}, we include CSM densities measured for other SN subtypes for context. We also show the epochs when the transition roughly occurred for all four SNe and corresponding mass-loss rates pre- and post-transition in Table \ref{tab:Comp_4SNe}. SN 2004dk and 2014C showed increases by at least a factor of five in mass-loss rate following the transition to type IIn-like optical spectra, while SN 2019oys is unconstrained before the transition (and has only one estimate of its mass-loss rate at $\sim 10^{-3}\rm M_{\odot}yr^{-1}$ post-transition \citep{Sfaradi_2024}). SN 2004dk only began showing signs of shock-driven hydrogen emission in its spectra at very late times (at 4684 days \citet{Mauerhan_2018}), and the date of transition to interaction is unclear (the most recent spectrum prior to the one at 4684 days was taken at 342 days). The emission in SN~2004dk was interpreted as coming from a wind-bubble scenario, where a fast wind excavates a lower-density cavity in the environment of the SN progenitor \citep{Pooley_2019}. The inferred jump in mass-loss rate is also only by a factor of 6 \citep{Pooley_2019}, lower than what was seen in SN~2014C where the inferred jump is by a factor $>$100 from late-time X-ray measurements \citep{Brethauer_2022}.

SN 2014C clearly showed a new regime of CSM at much higher density from the double peak in the radio lightcurve and later X-ray observations (see Figures \ref{fig:Radio_LCs} and \ref{fig:Xrays}). SN~2019oys seems to be an intermediate case between SN~2019yvr and SN~2014C, where the radio lightcurve did not have a double peak indicating a secondary shell, but rather had consistent interaction with high-density material revealed from a long-rising lightcurve at 15.5 GHz \citep{Sfaradi_2024}. 

In SN~2019yvr, we have shown that the object was clearly interacting with CSM before and after the hydrogen lines appeared (i.e. at 52 vs 287 days -- see Table \ref{tab:SSAindepoch} -- with the transition occurring at $\sim$ 100 days), but that the inferred mass-loss rate slightly increased following the optical appearance of shock-driven hydrogen lines. It is possible that SN 2019yvr has begun to show interaction with CSM from slightly enhanced mass loss which will give way to even denser material as in SN~2004dk. The key discrepancy between SN~2019yvr and SN~2014C (and potentially SNe~2004dk/2019oys) is that SN~2014C clearly only showed shock-driven hydrogen features after a large jump in CSM density (by a factor $>$ 100), but SN~2019yvr did not show an actual increase in CSM density at larger CSM radii (only a slight flattening of the density profile). It is important to note that these four objects (and SNe 2019oys and 2014C in particular) seem to represent an extreme edge of SNe Ib -- the majority of SNe Ib ($\sim$ 70--90 $\%$; see \cite{Margutti_2017}) show limited, if any, circumstellar interaction in the years following the explosion. 

\subsection{Multiwavelength interpretation}
The magnitude of the mass-loss rate we derive at radio and X-ray wavelengths is roughly consistent, assuming $\epsilon_{B}=0.01$,  with optical estimates from modeling of the H$\alpha$ profile of $(3-7) \times 10^{-5} \rm M_{\odot}\rm yr^{-1}$ \citep{Ferrari_2024}. The X-ray detection with \textit{Chandra} also implies a roughly similar mass-loss  magnitude at early times, suggesting a relatively homogeneous CSM (see Figure \ref{fig:ML}). We consider different scenarios that could explain the multiwavelength CSM picture.
\subsubsection{Insights into the mass-loss mechanism}
We speculate that the lack of hydrogen in the optical spectra at early times and presence at later times in SN~2019yvr could be caused by some combined optical depth or geometric effect where the hydrogen was hidden for the first $\sim$ 100 days, potentially due to being behind the optical photosphere and initially swept up by spherical ejecta overtaking an asymmetrically-distributed CSM, as was seen in iPTF14hls \citep{Andrews_2018}. However, the H$\alpha$ photosphere should be far below the CSM radius calculated from fits at 52 days (see Table \ref{tab:SSAindepoch}). Assuming an ejecta speed $\sim$ 10,000 km~s$^{-1}$ from optical measurements \citep{Ferrari_2024}, we obtain a photospheric radius of $\sim$ 5 $\times 10^{15}$ cm, a factor of two below the CSM radius derived at this epoch (see Table \ref{tab:SSAindepoch}).

 We note that our modeling has largely assumed spherical symmetry, which would not be the case if the CSM was created from binary mass transfer i.e. Roche-lobe Overflow \citep{Ercolino_2026}. In this case, the CSM would be distributed in a torus \citep{Mandal_2026}. However, any deviations from spherical symmetry would not affect our conclusion, namely that there was no significant increase in the inferred mass-loss rate when the H$\alpha$ emission appeared in optical spectra. There is no evidence for any change in the CSM geometry pre- and post- transition, making any correction for the geometry apply equally to both the pre- and post- transition derived quantities. In particular, the lack of any dramatic evolution in the radio light curves (see Figure \ref{fig:Radio_LCs}) suggests there was no significant change in the geometry of the radio-emitting region.

We postulate that another scenario that could explain the multiwavelength picture of SN~2019yvr is a situation where the nearby CSM is H-poor and the outer CSM is H-rich. In this scenario, the inner CSM would presumably be helium-rich, but still at lower density than typical SNe Ibn, e.g., SN 2023fyq \citep{RBW_2025}, and thus not producing strong narrow emission features. For a CSM speed of 100 km/s, the transition from hydrogen-rich mass-loss to hydrogen-poor mass-loss would have occurred $\sim$ 100 years pre-explosion (if the hydrogen-rich CSM was seen at $\sim$ 100 days for a shock-to-CSM speed ratio of 100-200).

This would mean that the progenitor likely lost most of its hydrogen envelope thousands of years before the explosion, and then lost the remaining amount of hydrogen from winds or weakened binary interaction at the $10^{-5}\,\mathrm{M_{\odot}
yr^{-1}}$ rate we infer from radio and X-ray data. The progenitor would then finally start losing some part of its helium layer at a similar mass-loss rate for $\sim$ 100 years preceding the final explosion. The rate we measure is consistent with expected mass-loss rates for a post-binary-interaction system that has lost its hydrogen envelope \citep{Ouchi_2017}. While this scenario does require some fine-tuning, the fact that objects like SN~2019yvr appear highly uncommon makes it more feasible. Further optical analysis of the helium and hydrogen features in particular  (as perhaps the helium absorption trough prevents definitive identification of the hydrogen emission at early times), and future radio and X-ray data probing further years pre-explosion would help clarify the likelihood of this proposed scenario.

 On a related note, another possibility, given especially the suggested supergiant companion of SN~2019yvr \citep{Sun_2021}, is that there was ejecta-companion interaction that produced hydrogen emission lines from unbound material from the companion (as explored for other supernovae, e.g., \cite{Hirai_2023}). This ejecta-companion interaction would be initially hidden by the photosphere in the supernova ejecta given the expected binary separation $\sim$ 10-1000 $R_{\odot}$ \citep{Hirai_2023}, and then revealed at later epochs. The CSM that produced radio and X-ray emission would then not necessarily be the material generating the hydrogen spectral features, and would be much further out than the ejecta-companion interaction region (which would maximally be at a scale of $\sim 10^{14}$ cm given expectations for binary systems \citep{Ercolino_2026}). Further modeling of this possibility is encouraged.

One final possibility is that the steady mass-loss rate enhancement by factors $\sim$ 2-5 starting at $\sim$ 100 days post-explosion caused the sudden appearance of shock-driven features in the optical spectrum. This enhancement could also suggest that the mass-loss mechanism changed (i.e., from binary to wind-driven) at the corresponding pre-explosion epoch ($\sim$ 100 years pre-explosion as explained). This scenario is certainly feasible, particularly as the detailed mechanisms of emission line formation due to interaction are still being studied and the density threshold for such features to appear is not well-constrained (for work on this question in type II supernovae, which may be slightly different, see \cite{Dessart_2023,WJG_2023}). We note, however, that it would be unusual for such a relatively minor excess over a wind-like profile to directly reveal such strong shock-driven hydrogen features. Further optical analysis and radiative transfer modeling of narrow emission features will help address this possibility.
\subsubsection{Insights into the progenitor system}
While \citet{Kilpatrick_2021} found that the pre-explosion source at the supernova location was consistent with a cool and inflated progenitor, \citet{Sun_2021} performed a re-analysis, including a detailed look at the star formation history in the SN environment, finding evidence that the SN progenitor was part of the oldest episode of star formation and thus may have had a relatively low ZAMS mass of $10.4_{-1.3}^{+1.5} \rm M_{\odot}$ \citep{Sun_2021}. They also suggested that the SED of the source found can be reproduced with a combined compact progenitor and extended companion, with the companion dominating the SED. The radio evolution we presented in this work shows no direct signs yet of a binary companion i.e., periodic behavior or a spectral inversion \citep{Ryder_2004,Chandra_2020,Maeda_2023}. However, as already mentioned, the high radio-derived shock speed at $>$ 30,000 km/s at 52 days post-explosion (see Table \ref{tab:SSAindepoch}) does also suggest a compact progenitor star \citep{Chevalier_2010}. 

Our results neither rule out nor confirm the suggested $< 2.6$ years pre-explosion hydrogen ejection from the progenitor proposed by \cite{Kilpatrick_2021}. Given that our radio and X-ray dataset start at 18 days post-explosion, if the shock to CSM speed ratio $>>$100 as suggested by the early-time shock speed measured $\sim$ 30,000 km/s, then the earliest pre-explosion timeframe we constrain is at least $\sim$ 5-10 years pre-explosion (this timeframe depends on the uncertain $\rm v_{CSM}$) from the first X-ray \textit{Swift} limit. However, the lack of strong H$\rm \alpha$ in early optical spectra would seem to rule out any possibility of a pre-explosion hydrogen ejection directly before the explosion.

\begin{deluxetable}{cccc}
\tablecaption{Comparison of SNe $\rm Ib \rightarrow IIn$ characteristics}
\tablehead{
\colhead{Supernova} & \colhead{Date of Transition ($\rm T_{tr}$) (days post-explosion)} & \colhead{Initial $\rm \dot{M}/v_{CSM}$} & \colhead{Post-transition $\rm \dot{M}/v_{CSM}$ }
}
\startdata
SN 2004dk & $342<\rm T_{tr}<4684$ & $\sim 6\times 10^{-9}$ & N/A \\
SN 2014C & $27<\rm T_{tr}<113$ & $\sim 3\times 10^{-7}$ & $\sim 8\times 10^{-4}$ \\
SN 2019oys & $1<\rm T_{tr}<150$  & N/A &$\sim 2\times 10^{-5}$ \\
SN 2019yvr & $79<\rm T_{tr}<118$  & $\sim 10^{-7}$ &  $\sim3\times  10^{-7}$  \\
\enddata
\tablecomments{ $\rm \dot{M}/v_{CSM}$ in standard units of ($\rm M_{\odot}yr^{-1}/km\, s^{-1}$). The transition is explicitly defined as when H$\alpha$ was first seen in optical spectra. The lower limits on the days since discovery when the transition occurred for SN~2004dk is taken from the last Wiserep \citep{Yaron_2012} spectrum that clearly showed no H$\alpha$ before the later-time spectrum at $>$ 4000 days presented in \cite{Mauerhan_2018}. The days for 2019yvr are from \cite{Ferrari_2024}, for 2014C from \cite{Danny_2014C} and for 2019oys from \cite{Sollerman_2020}. Date of explosion for SN~2014C is from \cite{Zhai_2025}. N/A is used when no measurement was made for a certain object. }
\end{deluxetable}
\label{tab:Comp_4SNe}

We emphasize again that it is particularly interesting that SN~2019yvr appeared optically similar to SN 2014C but did not show the same magnitude in the jump in density seen in SN 2014C (see Table \ref{tab:Comp_4SNe} and Figure \ref{fig:CSM_rhocomp} for a direct comparison of the four SNe $\rm Ib \rightarrow IIn$). The lack of evidence for a significant change (factor $>$ 10) in CSM density suggests that a scenario where the hydrogen is stripped at a very high rate in a potential common-envelope phase (within $\sim$ 1000 years pre-explosion) may be even rarer than initially thought from estimates using SNe 2014C and 2019yvr.

\section{Conclusion}
\label{sec:Conc}

In this work, we present the results of five years of radio (GMRT and VLA) and X-ray (\textit{Chandra} and \textit{Swift-XRT}) monitoring of SN 2019yvr. The radio light curves reveal the rise and fall of synchrotron emission across frequencies due to interaction with the CSM. The insights provided are:
\begin{enumerate}
    \item SN 2019yvr generated radio emission that is best-fit by a synchrotron self-absorbed model from ejecta interacting with CSM created by slightly decreasing mass-loss leading up to the explosion at $\sim 10^{-5}\,\mathrm{M_{\odot}yr^{-1}}$ (for an assumed CSM velocity of 100 km/s and $\epsilon_{B}$=0.01). The best-fit density values across epochs imply a CSM density profile $\rho \propto r^{-1.54\pm 0.31}$. The $\mathrm{\dot{M}}$ is of the same order of magnitude as optical estimates after the supernova transitioned to show shock-driven hydrogen lines at $\sim$ 100 days post-explosion. Furthermore, the radio SSA fits suggest a high shock speed $>$ 30000 km/s consistent with a compact progenitor.
    \item SN 2019yvr does not show any significant uptick in CSM density or free-free absorption in the radio evolution after the optical transition occurred. Based on late-time radio detections, SN 2019yvr shows CSM at least out to $\sim$ $10^{17}$ cm, suggesting mass was lost at a slightly decreasing rate $\sim 10^{-5}\,\mathrm{M_{\odot}yr^{-1}}$ for hundreds of years leading up to the explosion (assuming a shock to CSM speed ratio of $\sim$ 
    100--300).
    \item SN 2019yvr shows no evidence for a sudden increase in X-ray emission to deep limits, unlike SN 2014C. SN 2019yvr is weakly detected with \textit{Chandra} at 42 days post-explosion, providing an uncertain estimate of the mass-loss rate of $\sim$ $ 10^{-4}\,\mathrm{M_{\odot}yr^{-1}}$(assuming the same 100 km/s CSM speed). 
    \item The mass loss rate inferred is consistent with both binary stripping of a compact progenitor by an extended companion or elevated mass loss from a compact Wolf-Rayet star. However, we emphasize that the derived density profile suggests the mass-loss mechanism cannot be fully wind-like.
\end{enumerate} 
SN 2019yvr shows the great value in obtaining high-cadence radio and X-ray observations of stripped-envelope supernovae.
SN 2019yvr presents a new conundrum in the search for stripped envelopes: it seems that the characteristic emission lines from the interaction of the SN ejecta with the surrounding material can simply be missing in early optical spectra and seen in late-time spectra without drastic changes in CSM density. We suggest further analysis and multi-wavelength 3D radiative transfer modeling to understand the geometry of the configuration and the conditions needed for narrow emission line formation - indeed there are cases where there are no emission lines observed even in systems with high densities and mass loss rates (e.g. Ca-strong Transients-see \cite{Kumar_2026}). 

Furthermore, high-sensitivity and high-cadence radio and X-ray follow-up at early and late times of more stripped-envelope supernovae is needed to better understand the CSM environments of these stripped stars in general. Future radio facilities such as the Square Kilometer Array (SKA), which is poised to revolutionize SN radio studies \citep{Chandra_2026}, and the Deep Synoptic Array (DSA) \citep{Mclaughlin_2024}, will greatly increase the sample size of radio-observed core-collapse supernovae and provide great insight into mass-loss histories and progenitors across supernova subtypes in a systematic way.

\section{Acknowledgements}

We acknowledge insightful discussions with Makoto Johnstone, and appreciate the VLA data taken of NGC 4666 by Jeff Mangum and collaborators. RBW is supported by the National Science Foundation Graduate Research Fellowship Program under Grant number 2234693 and acknowledges support from the Virginia Space Grant Consortium.
PC acknowledges the support of this work provided by the National Aeronautics and Space Administration through Chandra Award Number GO4-25044X, along with GO3-24056X. 
M.M. and the METAL group at UVA acknowledge support in part from ADAP program grant No. 80NSSC22K0486, from the NSF grant AST-2206657, from the HST Grant GO-17596 by Space Telescope Science Institute, which is operated by the Association of Universities for Research in Astronomy, Inc., under NASA contract NAS5-26555,  and from the NSF under Cooperative Agreement 2421782 and the Simons Foundation grant MPS-AI-00010515 awarded to the NSF-Simons AI Institute for Cosmic Origins — CosmicAI, https://www.cosmicai.org/.
K.M. acknowledges support from JSPS KAKENHI grant (JP24KK0070, JP24H01810, and JP23H04894). CDK gratefully acknowledges support from the NSF through AST-2432037, the HST Guest Observer Program
through HST-SNAP-17070 and HST-GO-17706, and from JWST Archival Research through JWST-AR-6241 and
JWST-AR-5441.
Parts of this research were supported by the Australian Research Council Centre of Excellence for Gravitational Wave Discovery (OzGrav), through project number CE230100016.
We thank the staff of the VLA and the GMRT, who made these observations possible. GMRT is run by the National Centre for Radio Astrophysics of the Tata Institute of Fundamental Research. This paper employs a list of  Chandra datasets, obtained by the Chandra X-ray Observatory, contained in the Chandra Data Collection ~\dataset[DOI: 10.25574/cdc.640]{https://doi.org/10.25574/cdc.640}. The National Radio Astronomy Observatory and Green Bank Observatory are facilities of the U.S. National
Science Foundation operated under cooperative agreement by Associated Universities, Inc."
\appendix
\renewcommand{\thetable}{A\arabic{table}}
\setcounter{table}{0}
\renewcommand{\thesection}{A}

\startlongtable
\begin{deluxetable*}{cccccc}
\tablecaption{Radio Observations of SN 2019yvr\label{tab:radio_obs}}
\tablewidth{0pt}
\tablehead{
\colhead{UTC Date} &
\colhead{Epoch (Days after explosion)} & 
\colhead{Telescope} & 
\colhead{Freq (GHz)} & 
\colhead{Flux Density (mJy)} & 
\colhead{Image RMS ($\mu$Jy/Beam)}
}
\startdata
14 January 2020 & 23   & GMRT & 1.25 & $<$0.3               & 15 \\
12 February 2020 & 52   & VLA-C  & 3    & 4.86 $\pm$ 0.32      & 70 \\
15 February 2020 & 55   & GMRT & 0.65 & $<3.6$    & 30 \\
22 February 2020 & 62   & GMRT & 1.25 & 0.80 $\pm$ 0.20      & 100 \\
13 March 2020    & 82   & VLA-C  & 3    & 5.61 $\pm$ 0.35      & 82 \\
3 July 2020      & 194  & VLA-B  & 9    & 0.654 $\pm$ 0.015    & 8 \\
7 July 2020      & 198  & VLA-B  & 11   & 0.525 $\pm$ 0.012    & 6 \\
7 July 2020      & 198  & VLA-B  & 1.25 & 3.6 $\pm$ 0.2        & 240 \\
7 July 2020      & 198  & VLA-B  & 1.5  & 3.22 $\pm$ 0.25      & 125 \\
7 July 2020      & 198  & VLA-B  & 3    & 2.4 $\pm$ 0.3        & 60 \\
7 July 2020      & 198  & VLA-B  & 5    & 1.335 $\pm$ 0.074    & 40 \\
7 July 2020      & 198  & VLA-B  & 7    & 0.755 $\pm$ 0.07     & 20 \\
10 August 2020   & 232  & GMRT & 0.65 & 1.93 $\pm$ 0.11      & 50 \\
18 August 2020   & 240  & GMRT & 1.25 & 2.93 $\pm$ 0.39      & 40 \\
4 October 2020   & 287  & VLA -B & 1.25 & 5.23 $\pm$ 0.46      & 200 \\
4 October 2020   & 287  & VLA-B  & 1.5  & 4.04 $\pm$ 0.15      & 120 \\
4 October 2020   & 287  & VLA-B & 3    & 2.44 $\pm$ 0.12      & 30 \\
4 October 2020   & 287  & VLA-B  & 5    & 1.364 $\pm$ 0.025    & 20 \\
4 October 2020   & 287 & VLA-B  & 7    & 0.963 $\pm$ 0.029    & 50 \\
4 October 2020   & 287  & VLA-B  & 9    & 0.71 $\pm$ 0.022     & 45 \\
4 October 2020   & 287  & VLA-B  & 11   & 0.61 $\pm$ 0.019     & 10 \\
31 December 2020 & 375  & GMRT & 1.25 & 3.05 $\pm$ 0.13      & 60 \\
1 January 2021   & 376  & GMRT & 0.75 & 2.79 $\pm$ 0.11      & 150 \\
13 February 2021 & 419  & VLA-A  & 1.25 & 2.48 $\pm$ 0.12      & 70 \\
13 February 2021 & 419  & VLA-A  & 1.5  & 2.224 $\pm$ 0.061    & 200 \\
13 February 2021 & 419  & VLA-A  & 3    & 1.235 $\pm$ 0.014    & 12 \\
13 February 2021 & 419  & VLA-A  & 5    & 0.769 $\pm$ 0.005    & 6.5 \\
13 February 2021 & 419  & VLA-A & 7    & 0.525 $\pm$ 0.010    & 6 \\
14 September 2021& 632  & GMRT & 1.25 & 1.39 $\pm$ 0.13      & 60 \\
12 December 2021 & 721  & VLA-B & 3    & 0.414 $\pm$ 0.139    & 30 \\
24 January 2022  & 764  & GMRT & 0.75 & 1.97 $\pm$ 0.13      & 30 \\
25 January 2022  & 765  & GMRT & 1.25 & 0.747 $\pm$ 0.048    & 40 \\
20 May 2022      & 880  & VLA-A  & 1.25 & 0.596 $\pm$ 0.024    & 50 \\
20 May 2022      & 880  & VLA-A  & 1.75 & 0.537 $\pm$ 0.016    & 80 \\
20 May 2022      & 880  & VLA-A  & 3    & 0.253 $\pm$ 0.013    & 80 \\
20 May 2022      & 880  & VLA-A  & 5    & 0.188 $\pm$ 0.006    & 15 \\
20 May 2022      & 880  & VLA-A  & 7    & 0.125 $\pm$ 0.008    & 10 \\
8 June 2022      & 900  & GMRT & 1.25 & 0.517 $\pm$ 0.16     & 50 \\
27 November 2022 & 1071 & VLA-C  & 1.5  & $<$45                & 200 \\
27 November 2022 & 1071 & VLA-C  & 3    & $<$1.2               & 400 \\
27 November 2022 & 1071 & VLA-C  & 5    & $<$1.5               & 50 \\
27 November 2022 & 1071 & VLA-C  & 7    & $<$0.6               & 30 \\
5 May 2023       & 1230 & GMRT & 0.4  & $<$1.8               & 170 \\
6 May 2023       & 1231 & GMRT & 1.25 & $<$0.3               & 80 \\
7 May 2023       & 1232 & GMRT & 0.75 & $<$2.25              & 500 \\
18 March 2024    & 1548 & GMRT & 0.4  & $<$2.55              & 300 \\
20 March 2024    & 1550 & GMRT & 0.75 & $<$0.6               & 200 \\
22 March 2024    & 1552 & GMRT & 1.25 & $<$1.35              & 50 \\
11 May 2024      & 1602 & GMRT & 0.4  & $<$9.9               & 800 \\
1 June 2024      & 1623 & GMRT & 1.25 & $<$0.75              & 50 \\
29 July 2024     & 1681 & GMRT & 0.75 & $<$0.72              & 45 \\
2 November 2024  & 1777 & VLA-B  & 4.93 & 0.094 $\pm$ 0.011    & 15 \\
11 November 2024 & 1786 & GMRT & 0.4  & $<$2.25              & 300 \\
10 November 2024 & 1785 & GMRT & 0.75 & $<$1.05              & 450 \\
9 November 2024  & 1784 & GMRT & 1.25 & $<$1.05              & 75 \\
\enddata
\tablecomments{All upper limits are 3$\sigma$. VLA configuration is given in the Telescope row. The date of explosion is taken as December 22, 2019 from \cite{Ferrari_2024}}
\end{deluxetable*}
%

\facilities{Swift(XRT), Chandra, VLA, GMRT}



\let\cleardoublepage\clearpage

\

\bibliography{ms}{}
\bibliographystyle{aasjournal}



\end{document}